\documentclass[fleqn,usenatbib]{mnras}

\usepackage{newtxtext,newtxmath}
\usepackage[T1]{fontenc}
\usepackage{ae,aecompl}

\usepackage{times}
\usepackage{epsfig}
\usepackage{ifpdf}
\usepackage{graphicx}
\usepackage{verbatim} 
\usepackage{amsmath}
\usepackage{amssymb}
\usepackage{calc}
\usepackage[normal]{threeparttable}
\usepackage{longtable}
\usepackage{pdflscape}
\newcommand{\comments}[1]{} 
\usepackage{placeins} 
\usepackage{float}
\usepackage{soul}
\graphicspath{{./plots/}}
\title[HI absorption survey of GAMA\,23]{FLASH Early Science -- Discovery of an intervening HI 21-cm absorber from an ASKAP survey of the GAMA\,23 field}

\author[J.~R. Allison et al.]{
J.~R. Allison,$^{1,2}$\thanks{E-mail: james.allison@physics.ox.ac.uk} E.~M. Sadler,$^{3,4,2}$ S. Bellstedt,$^{5}$ L.~J.~M. Davies,$^{5}$ S.~P. Driver,$^{5,6}$ \newauthor S.~L. Ellison,$^{7}$ M. Huynh,$^{5,8}$ A.~D. Kapi\'nska,$^{9}$ E.~K. Mahony,$^{4}$ V. A. Moss,$^{4,3}$ \newauthor A. S. G. Robotham,$^{5}$ M.~T. Whiting,$^{4}$  S.~J. Curran,$^{10}$ J. Darling$^{11}$, A.~W. Hotan,$^{4}$ \newauthor R.~W. Hunstead,$^{3}$\thanks{Deceased} B.~S. Koribalski,$^{4}$ C.~D.~P. Lagos,$^{2,5}$ M. Pettini,$^{12}$  K.~A. Pimbblet,$^{13}$ \newauthor and M.~A. Voronkov$^{4}$
\\
$^{1}$Sub-Dept. of Astrophysics, Department of Physics, University of Oxford, Denys Wilkinson Building, Keble Rd., Oxford, OX1 3RH, UK\\
$^{2}$ARC Centre of Excellence for All-sky Astrophysics in 3 Dimensions (ASTRO 3D)\\
$^{3}$Sydney Institute for Astronomy, School of Physics A28, University of Sydney, Sydney, NSW 2006, Australia\\
$^{4}$CSIRO Astronomy and Space Science, PO Box 76, Epping, NSW 1710, Australia\\
$^{5}$International Centre for Radio Astronomy Research (ICRAR), University of Western Australia, Crawley, WA 6009, Australia \\
$^{6}$School of Physics \& Astronomy, University of St Andrews, North Haugh, St Andrews, KY16 9SS, UK, \\
$^{7}$Department of Physics \& Astronomy, University of Victoria, Finnerty Road, Victoria, British Columbia, V8P 1A1, Canada\\
$^{8}$CSIRO Astronomy and Space Science, 26 Dick Perry Avenue, Kensington WA 6151, Australia\\
$^{9}$National Radio Astronomy Observatory, 1003 Lopezville Rd., Socorro NM 87801, USA\\
$^{10}$School of Chemical and Physical Sciences, Victoria University of Wellington, PO Box 600, Wellington 6140, New Zealand\\
$^{11}$Department of Astrophysical and Planetary Sciences, University of Colorado, 389 UCB, Boulder, CO 80309-0389, USA\\
$^{12}$Institute of Astronomy, University of Cambridge, Madingley Road, Cambridge, CB3 0HA, UK\\
$^{13}$E. A. Milne Centre for Astrophysics, University of Hull, Cottingham Road, Kingston-upon-Hull, HU6 7RX, UK\\
}

\date{Accepted XXX. Received YYY; in original form ZZZ}

\pubyear{2020}

\begin{document}
\label{firstpage}
\pagerange{\pageref{firstpage}--\pageref{lastpage}}
\maketitle

\begin{abstract}
We present early science results from the First Large Absorption Survey in \mbox{H\,{\sc i}} (FLASH), a spectroscopically blind survey for 21-cm absorption lines in cold hydrogen (\mbox{H\,{\sc i}}) gas at cosmological distances using the Australian Square Kilometre Array Pathfinder (ASKAP). We have searched for \mbox{H\,{\sc i}} absorption towards 1253 radio sources in the GAMA\,23 field, covering redshifts between $z = 0.34$ and $0.79$ over a sky area of approximately 50\,$\mathrm{deg}^2$. In a purely blind search we did not obtain any detections of 21-cm absorbers above our reliability threshold. Assuming a fiducial value for the \mbox{H\,{\sc i}} spin temperature of $T_{\rm spin}$ = 100\,K and source covering fraction $c_{\rm f} = 1$, the total comoving absorption path length sensitive to all Damped Lyman $\alpha$ Absorbers (DLAs; $N_{\rm HI} \geq 2 \times 10^{20}$\,cm$^{-2}$) is $\Delta{X} = 6.6 \pm 0.3$ ($\Delta{z} = 3.7 \pm 0.2$) and super-DLAs ($N_{\rm HI} \geq 2 \times 10^{21}$\,cm$^{-2}$) is $\Delta{X} = 111 \pm 6$ ($\Delta{z} = 63 \pm 3$). We estimate upper limits on the \mbox{H\,{\sc i}} column density frequency distribution function that are consistent with measurements from prior surveys for redshifted optical DLAs, and nearby 21-cm emission and absorption. By cross-matching our sample of radio sources with optical spectroscopic identifications of galaxies in the GAMA\,23 field, we were able to detect 21-cm absorption at $z = 0.3562$ towards NVSS\,J224500$-$343030, with a column density of $N_{\rm HI} = (1.2 \pm 0.1) \times 10^{20}\,(T_{\rm spin}/100\,\mathrm{K})$\,cm$^{-2}$. The absorber is associated with GAMA\,J22450.05$-$343031.7, a massive early-type galaxy at an impact parameter of 17\,kpc with respect to the radio source and which may contain a massive ($M_{\rm HI} \gtrsim 3 \times 10^{9}$\,M$_{\odot}$) gas disc. Such gas-rich early types are rare, but have been detected in the nearby Universe.
\end{abstract}


%%%%%%%%%%%%%%%%%%%%%%%%%%%%%%%%%%%%%%%%%%%%%%%%%%

%%%%%%%%%%%%%%%%% BODY OF PAPER %%%%%%%%%%%%%%%%%%

\section{Introduction}

Star formation and supermassive black hole (SMBH) growth are two important processes in galaxies that influence their evolution throughout cosmic history. However, we do not yet understand why the global rates of star formation (e.g. \citealt{Hopkins:2006b, Madau:2014, Driver:2018}) and SMBH growth (e.g. \citealt{Ueda:2003, Shankar:2009}) both peaked at $z \approx 2$, and then declined by an order magnitude to this epoch. It is clear that a ready supply of cold ($T_{\rm k} \ll 10^{4}$\,K) gas is important; in the nearby Universe the surface densities of star formation and neutral gas, particularly the molecular component, are strongly correlated (e.g. \citealt{Schmidt:1959, Kennicutt:1998, Bigiel:2008}), and likewise radiatively efficient active galactic nuclei (AGNs) are predominantly hosted by star-forming galaxies with a central young stellar population and therefore ample cold gas reservoirs (e.g. \citealt{Kauffmann:2003, Kauffmann:2007, Kauffmann:2009, LaMassa:2013, Ellison:2019}). Determining how the neutral interstellar medium in galaxies has evolved over the history of the Universe is therefore a key component in understanding their evolution. 
Much of our knowledge of the global content of neutral gas in galaxies comes from observing hydrogen gas, the most common element in the Universe. In its neutral atomic (\mbox{H\,{\sc i}}) phase, hydrogen is readily detectable in nearby galaxies via 21-cm emission at radio wavelengths (see \citealt{Giovanelli:2016} for a review) or, at cosmological distances, through Lyman $\alpha$ ($n = 1-2$) absorption in the ultraviolet and visible bands (see \citealt{Wolfe:2005}). The total \mbox{H\,{\sc i}} mass density shows comparatively less evolution over cosmological time-scales than that of star formation, decreasing by at most two-fold since $z \approx 2$ (e.g. \citealt{Zwaan:2005, Martin:2010, Braun:2012, Noterdaeme:2012, Zafar:2013, Crighton:2015, Sanchez-Ramirez:2016, Bird:2017, Rhee:2018}), suggesting that much of the neutral gas content of galaxies is replenished over these time-scales.
	
In contrast, observations of the molecular gas at cosmological distances, traced by emission lines from the low-$J$ rotational transitions of carbon monoxide ($^{12}$C$^{16}$O), suggest a much stronger evolution of the coldest ($T_{\rm k} \sim 10$\,K) gas (see \citealt{Carilli:2013} for a review). Samples of star-forming and ultraluminous galaxies (e.g. \citealt{Tacconi:2013, Combes:2013, Magdis:2014, Villaneuva:2017, Isbell:2018}) have revealed a decrease in the molecular gas fraction in these galaxies since the peak of star formation at $z \approx 2$. Recent deep observations for CO emission lines in the \emph{Hubble} Ultra Deep Field with the Atacama Large Millimetre Array (ALMA) also show a decrease in the knee of the CO luminosity function over the same period, implying a corresponding decrease in the cosmological density of H$_{2}$ that appears to closely match that of the star formation rate (\citealt{Decarli:2016b, Decarli:2019}). 
	
Radio surveys for the \mbox{H\,\sc i} 21-cm absorption line, seen in the spectra of radio sources, afford us an important additional tool in establishing a census of the neutral gas in galaxies at cosmological distances (\citealt{Kanekar:2004, Morganti:2015}). 21-cm absorption lines are detected in individual galaxies at luminosity distances well beyond that currently obtainable for emission line surveys ($z \lesssim 0.4$; \citealt{Fernandez:2016}), and are only limited by the observable band of the telescope and a sufficiently large population of detectable background sources. At redshifts below $z = 1.7$, the ultraviolet Lyman $\alpha$ line is no longer redshifted into the optical window and can only be observed using the \emph{Hubble Space Telescope}. Searches for DLAs at these redshifts are therefore observationally expensive and necessarily lead to smaller sample sizes and correspondingly poorer statistical constraints on the cosmological evolution of \mbox{H\,{\sc i}} than at higher redshifts (e.g. \citealt{Neeleman:2016, Rao:2017}). To improve the DLA detection rate, it is common practice to select quasars that have existing Mg\,{\sc ii}\,$\lambda\lambda$\,2796, 2803\,\AA\ absorption with equivalent widths greater than about 0.5\,\AA\ (e.g. \citealt{Rao:2006, Ellison:2009, Rao:2017}). There is a concern that this might bias the identification of \mbox{H\,{\sc i}} absorbers, particularly against those with low column densities and metallicities (e.g. \citealt{Peroux:2004, Dessauges-Zavadsky:2009, Neeleman:2016, Berg:2017}, but see also \citealt{Rao:2017}). The 21-cm absorption line is therefore key to establishing the \mbox{H\,{\sc i}} content of galaxies at these intermediate cosmological distances. 

Importantly, the equivalent width of the 21-cm absorption line depends on both the \mbox{H\,{\sc i}} column density and its excitation (spin) temperature along the line of sight to the radio source. If the \mbox{H\,{\sc i}} column density can be determined independently via either 21-cm emission (e.g. \citealt{Reeves:2016, Borthakur:2016, Gupta:2018}) or Lyman $\alpha$ absorption (see \citealt{Kanekar:2014a} and references therein), then the spin temperature in the absorber can be inferred. However, in doing so one must be careful to consider the relative sizes of the foreground absorber and background continuum source, particularly at radio wavelengths where the source may be significantly larger than the spatial distribution of opaque \mbox{H\,{\sc i}} structures (see e.g. \citealt{Curran:2005, Braun:2012}). 

The spin temperature enables comparisons to be drawn with the multiphase neutral interstellar medium (ISM) seen in the Milky Way and Local Group galaxies (e.g. \citealt{Dickey:1994, Dickey:2000, Heiles:2003b, Roy:2013b, Murray:2018}), in particular the inferred fraction of cold ($T_{\rm k} \sim 100$\,K) neutral medium (CNM), which is the component of the atomic gas most likely to trace star formation. Searches for 21-cm absorption in known DLAs suggest that the typical spin temperature of high column density systems is anticorrelated with their metallicity and may increase at redshifts above $z = 2$, beyond the peak of star formation in the Universe (\citealt{Kanekar:2014a}). This would be consistent with a model whereby relatively metal-poor DLAs in the early Universe (e.g. \citealt{Rafelski:2012, Cooke:2015, DeCia:2018}) lacked sufficient coolants in the gas to form a significant fraction of CNM via fine structure line cooling. Recently, by explicitly modelling the source covering fraction as a function of angular diameter distance, \cite{Curran:2019a, Curran:2017a} used literature searches for redshifted 21-cm absorption to show that the spin temperature has evolved with the star formation rate history of the Universe. However, we caution that this uses an evolutionary model for the covering fraction of radio sources that would mimic any perceived evolution in the spin temperature and so future model-independent methods are required to verify such a claim.

Future large-scale, radio-selected 21-cm absorption line surveys will be able to statistically determine the cosmological evolution of the physical state of the neutral atomic gas at intermediate redshifts (e.g. \citealt{Darling:2011, Allison:2016b}). Several of these surveys will be undertaken using pathfinder telescopes to the planned Square Kilometre Array. These include the First Large Absorption Survey in \mbox{H\,{\sc i}} (FLASH; e.g. \citealt{Allison:2016b}) with the Australian Square Kilometre Pathfinder (ASKAP; \citealt{Johnston:2007}),  the South African MeerKAT Absorption Line Survey (MALS; \citealt{Gupta:2016}) and the Search for \mbox{H\,{\sc i}} Absorption with AperTIF (e.g. \citealt{Oosterloo:2009}). 

\begin{table*}
\caption{Summary of observations of the GAMA\,23 field using the lower frequency bands of ASKAP-12. Each observation was assigned a unique scheduling block identification (SBID) number and has two interleaved pointing centres labelled A and B. $t_{\rm obs}$ denotes the duration of the observation. $\sigma_{\rm chan}$ is the rms noise per 18.5\,kHz channel, where we give the interquartile range over the whole bandwidth and all 36 PAF beams. $\Delta{v}_{\rm chan}$ is the spectral resolution in rest-frame radial velocity and $z_{\rm HI}$ is the \mbox{H\,{\sc i}} 21-cm redshift range across the observed band.}\label{table:GAMA23_obs}
\centering
\begin{tabular*}{\textwidth}{llccccccccc}
\hline \hline
SBID & Date & \multicolumn{3}{c}{Pointing centre}  & Frequency band$^{a}$ & No. ant.$^{a}$ & $t_{\rm obs}$ & $\sigma_{\rm chan}$ & $\Delta{v}_{\rm chan}$ & $z_{\rm HI}$ \\
& UTC & & RA\,[J2000] & Dec.\,[J2000] & [MHz] & & [h] & [mJy\,beam$^{-1}$] & [km\,s$^{-1}$] & \\
\hline
2955 & 2016/12/17 & A & $23^{\rm{h}} 00^{\rm{m}} 00{\fs}000$ & $-32\degr 30\arcmin 00{\farcs}00$ & 864.5 -- 1056.5 & 12 & 4 & 12 -- 14 & 5.2 -- 6.4 & 0.34 -- 0.64 \\
& & B & $23^{\rm{h}} 03^{\rm{m}} 01{\fs}096$ & $-32\degr 29\arcmin 51{\farcs}89$ & & & & & & \\
2961 & 2016/12/18 & A & $23^{\rm{h}} 00^{\rm{m}} 00{\fs}000$ & $-32\degr 30\arcmin 00{\farcs}00$ & 864.5 -- 1056.5 & 12 & 4 & 12 -- 14 & 5.2 -- 6.4 & 0.34 -- 0.64 \\
& & B & $23^{\rm{h}} 03^{\rm{m}} 01{\fs}096$ & $-32\degr 29\arcmin 51{\farcs}89$ & & & & & & \\
4996 & 2018/01/13 & A & $22^{\rm{h}} 46^{\rm{m}} 13{\fs}600$ & $-32\degr 15\arcmin35{\farcs}27$ & 792.5 -- 1032.5 & 15 & 11 & 6.5 -- 8.0 & 5.4 -- 7.0 & 0.38 -- 0.79 \\
& & B & $22^{\rm{h}} 48^{\rm{m}} 21{\fs}955$ & $-32\degr 42\arcmin 31{\farcs} 25$ & & & & & & \\
5000 & 2018/01/15 & A & $23^{\rm{h}} 11^{\rm{m}} 46{\fs}399$ & $-32\degr 15\arcmin 35{\farcs}27$ & 792.5 -- 1032.5 & 15 & 11 & 6.8 -- 8.4 & 5.4 -- 7.0 & 0.38 -- 0.79 \\
& & B & $23^{\rm{h}} 13^{\rm{m}} 54{\fs}754$ & $-32\degr 42\arcmin 31{\farcs}25$ & & & & & & \\
5229 & 2018/03/16 & A & $22^{\rm{h}} 46^{\rm{m}} 13{\fs}600$ & $-32\degr 15\arcmin 35{\farcs}27$ & 816.5 -- 1056.5 & 16 & 11 & 5.4 -- 6.2 & 5.2 -- 6.8 & 0.34 -- 0.74 \\
& & B & $22^{\rm{h}} 48^{\rm{m}} 21{\fs}955$ & $-32\degr 42\arcmin 31{\farcs}25$ & & & & & & \\
5232 & 2018/03/18 & A & $23^{\rm{h}} 11^{\rm{m}} 46{\fs}399$ & $-32\degr 15\arcmin 35{\farcs}27$ & 816.5 -- 1056.5 & 16 & 11 & 6.6 -- 7.7 & 5.2 -- 6.8 & 0.34 -- 0.74 \\
& & B & $23^{\rm{h}} 13^{\rm{m}} 54{\fs}754$ & $-32\degr 42\arcmin 31{\farcs}25$ & & & & & & \\
\hline \hline
\end{tabular*}
\medskip
$^{a}$Note that the available frequency band and antennas changed throughout commissioning of the array and correlator.
\end{table*}

In this paper, we present early science results from ASKAP FLASH, a spectroscopically blind survey for \mbox{H\,{\sc i}} absorption which will eventually cover the entire southern sky ($\delta < +10\degr$) between $z = 0.4$ and $1.0$. Such a survey with ASKAP is enabled by combining a 30\,$\mathrm{deg}^2$ field of view, 300\,MHz of instantaneous bandwidth and a frequency band that below 1\,GHz is typically free of any radio frequency interference. Several recent results from observations of radio galaxies and quasars during ASKAP commissioning have demonstrated feasibility (e.g. \citealt{Allison:2015, Allison:2016a, Moss:2017, Allison:2017, Glowacki:2019, Allison:2019}). Here, we have used observations of the 23\,hr field of the Galaxy And Mass Assembly survey (GAMA; \citealt{Liske:2015})\footnote{\url{http://www.gama-survey.org}}, a 50\,$\mathrm{deg}^2$ area of the southern sky that contains spectroscopic information for galaxies to an $i$-band magnitude limit of $i < 19.2$\,mag. Due to commissioning constraints on the available correlator hardware, our observations covered \mbox{H\,{\sc i}} redshifts between $z_{\rm HI} = 0.34$ and 0.79, which is slightly different to that expected for FLASH. Our broad goal was to carry out the first wide-field spectroscopically blind search for 21-cm absorption at cosmological distances (see also \citealt{Darling:2004, Darling:2011}) and establish the methodology that will be employed in future, larger surveys.

We structure this paper as follows. In \autoref{section:observations_data}, we describe briefly our observations and data analysis, referring the interested reader to previous work and providing the salient updates. We discuss the results of our survey in \autoref{section:results}, both when spectroscopically blind and when we cross-match with optically identified galaxies. We summarize our conclusions in \autoref{section:summary}. In all distance calculations dependent on the cosmological parameters we adopt a flat lambda cold dark matter ($\Lambda$CDM) cosmology with $H_{0} = 70$\,km\,s$^{-1}$, $\Omega_{\rm m} = 0.3$ and $\Omega_{\Lambda} = 0.7$ (e.g. \citealt{Spergel:2007}). 
 
\begin{figure*}
\centering
\includegraphics[width=0.9\textwidth]{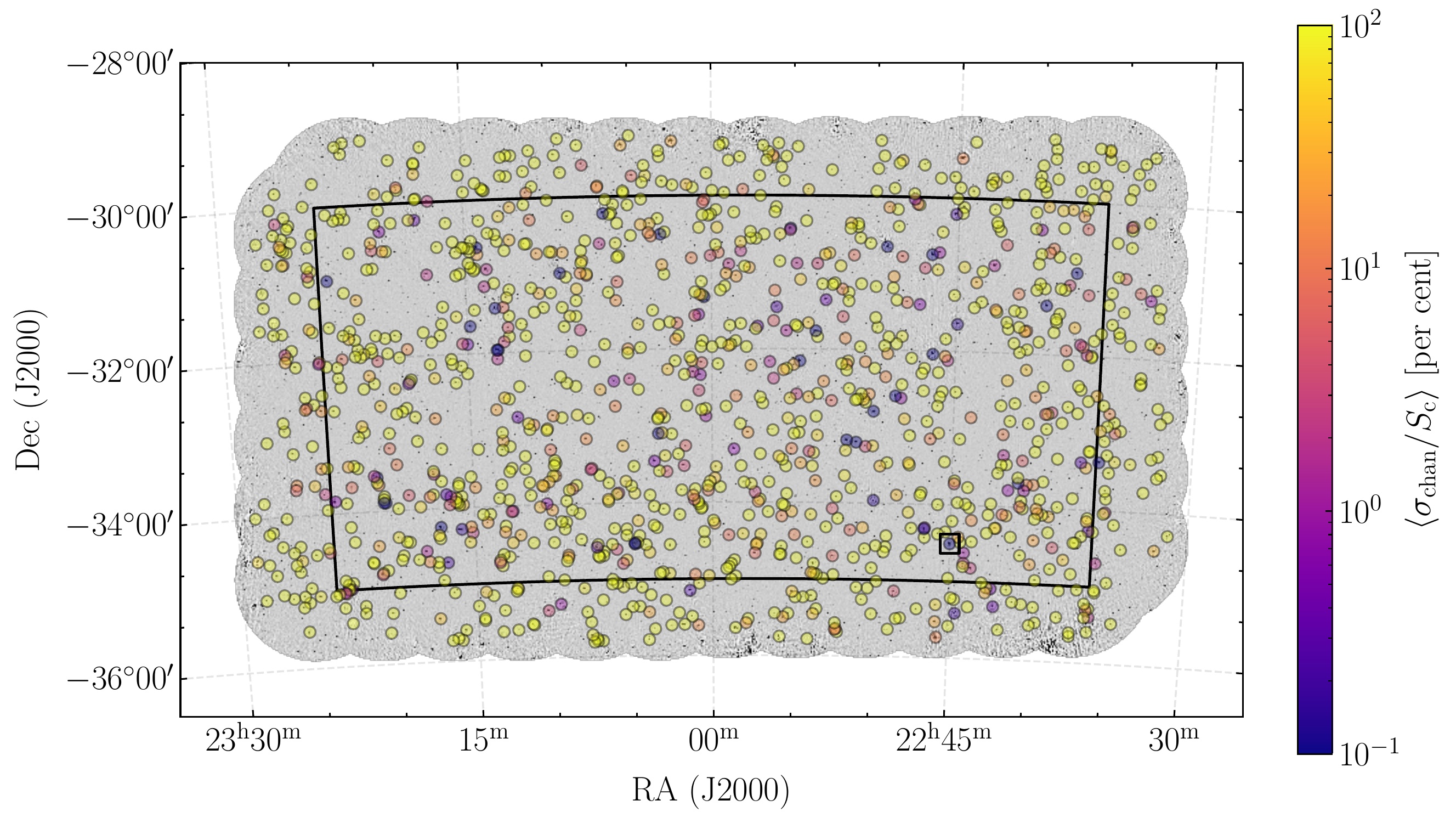}
\caption{An ASKAP-12 mosaic of the 900\,MHz radio continuum in the GAMA\,23 field (grey scale image). The small circles indicate the 1253 radio sources towards which we searched for \mbox{H\,{\sc i}} 21-cm absorption. The colour scale denotes the median rms noise per 18.5\,kHz channel as a fraction of the continuum; darker circles indicate higher sensitivity to 21-cm absorption. The small black square indicates the position of NVSS\,J224500$-$343030, towards which we have detected \mbox{H\,{\sc i}} 21-cm absorption (see text for details). The larger black boundary denotes the extent of the GAMA\,23 field (\citealt{Liske:2015}).}
\label{figure:GAMA23_mosaic}
\end{figure*}

\begin{figure*}
\centering
\includegraphics[width=0.975\textwidth]{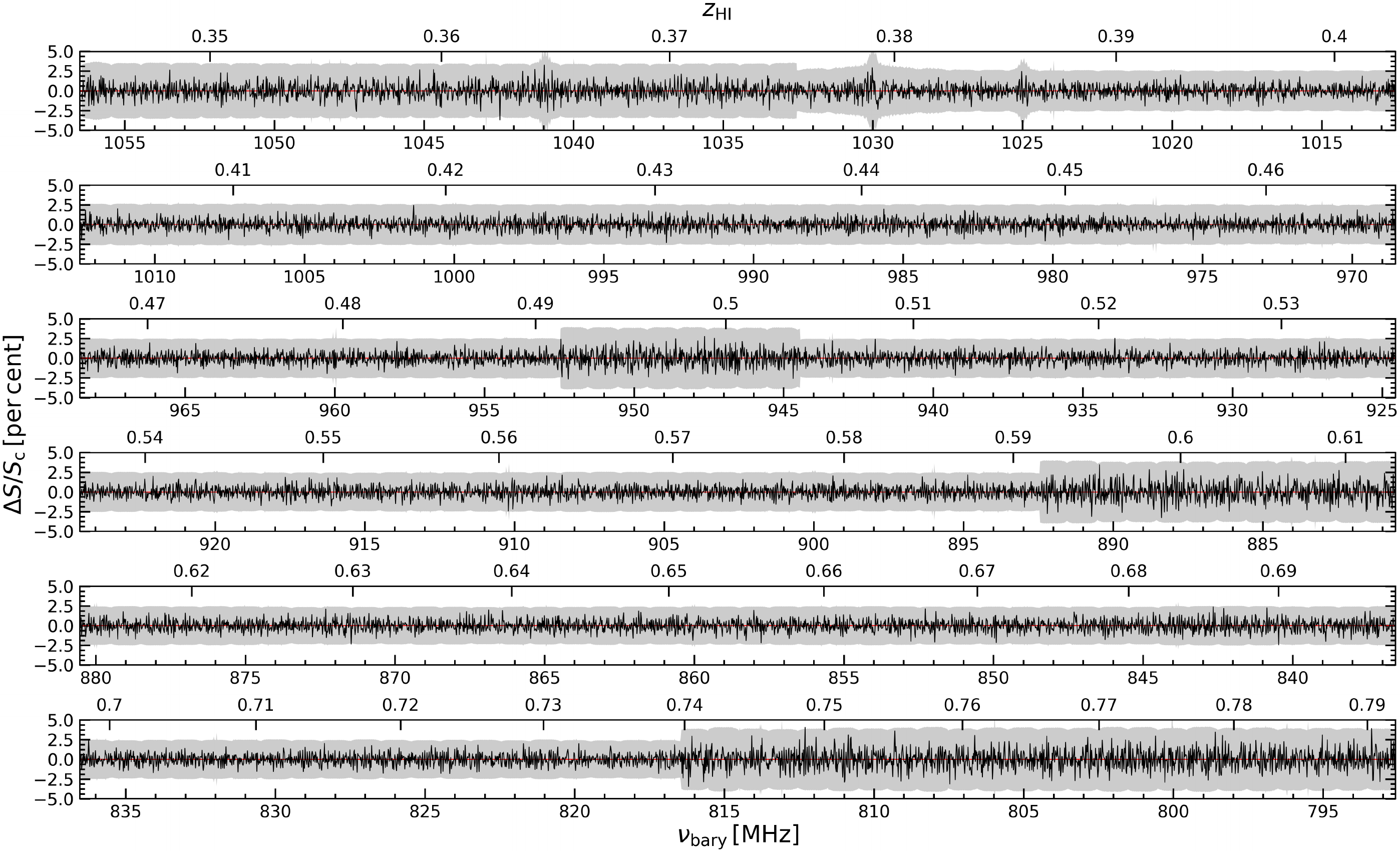}
\caption{An example ASKAP-12 spectrum towards the radio source NVSS\,J224500$-$343030 in the GAMA\,23 field. The continuum-subtracted data (black line) are given as a fraction of the continuum flux density. The grey region denotes 5 times the rms noise measured from the channel images. Variations in sensitivity across the band are dominated by observations of different duration at distinct frequencies (see \autoref{table:GAMA23_obs}), and occasional failures of individual correlator cards. On the lower horizontal axis is the solar barycentric-corrected observed frequency and the upper horizontal axis is the corresponding \mbox{H\,{\sc i}} redshift.}
\label{figure:GAMA23_example_spectrum}
\end{figure*}

\section{Observations and Data}\label{section:observations_data}

The GAMA\,23 field (\citealt{Liske:2015}) was observed with ASKAP as part of the early science programs of the Evolutionary Map of the Universe survey (EMU; \citealt{Norris:2011, Leahy:2019}) and the Deep Investigation of Neutral Gas Origins survey (DINGO)\footnote{\url{https://dingo-survey.org}}. These observations used ASKAP-12, a sub-array of ASKAP that had between 12 and 16 operational antennas and a bandwidth between 192 and 240\,MHz, primarily to carry out early science and commission the Mrk\,II Phased Array Feeds (PAFs; see e.g. \citealt{Chippendale:2015}). In order to closely match the frequency band we expect to use with the full FLASH survey (700 - 1000\,MHz), and to maximize the comoving path for intervening absorption detection, we only chose those observations that used the lowest available band at that time. Three distinct observations were undertaken from 2016 December to 2018 March; we summarize these in \autoref{table:GAMA23_obs}. The spectral resolution is set by the 18.5\,kHz channelization, which for the observed bands corresponds to rest-frame velocities in the range 5.2 to 7.0\,km\,s$^{-1}$. Maximum signal-to-noise PAF beams were electronically formed in a $6 \times 6$ square pattern on the sky with a separation of 0.9\,deg between adjacent beam centres. Each observation had two interleaved pointing positions that were switched every 15\,min so as to enable a uniform sensitivity pattern across the field and avoid correlated noise.

A dedicated \textsc{ASKAPsoft} data processing pipeline is currently being developed for the full FLASH survey, adapting the procedure discussed by \cite{Kleiner:2019}. However, to facilitate processing of the early science data presented here we have followed the procedure described by \cite{Allison:2015, Allison:2017, Allison:2019} and refer the interested reader to that work for further details. The ingested data from the correlator were written to measurement set format and then flagged for digital glitches and autocorrelations, averaged, and split, using the \textsc{CASA}\footnote{\url{http://www.casa.nrao.edu}} package (\citealt{McMullin:2007}). For each PAF beam, we split the data at full spectral resolution into 1-MHz band chunks, equal to the beam-forming intervals. Separately, a single 1MHz-averaged data set was produced to be used for high signal-to-noise continuum imaging and self-calibration. Further automated flagging, calibration and imaging of the data was carried out using the \textsc{MIRIAD}\footnote{\url{http://www.atnf.csiro.au/computing/software/miriad/}} package (\citealt{Sault:1995}). Short observations of PKS\,B1934$-$638 (approximately 5\,min per PAF beam) were used to obtain initial solutions for the complex antenna gains and to calibrate the flux scale (using the model of \citealt{Reynolds:1994}). We then used the catalogue of the NRAO Very Large Array Sky Survey (NVSS; \citealt{Condon:1998}) to construct a reference sky model for initial self-calibration, followed by iterative self-calibration using deconvolved models from the ASKAP data. Continuum images were formed using multifrequency synthesis with a Brigg's robustness weighting parameter of 0.5, giving an average synthesized beam of $40 \times 25$\,arcsec (full width at half maximum; FWHM) with a standard deviation of $10$\,per\,cent between the PAF beams and observations. At the median \mbox{H\,{\sc i}} 21-cm redshift of our observations ($z = 0.57$), this resolution corresponds to a physical scale of approximately $200$\,kpc. This is not sufficient to accurately determine the ratio of compact to extended radio emission for a given source, and so estimate the areal covering fraction of the \mbox{H\,{\sc i}} distributed in front of the source. The effect of the unknown source covering fraction is to underestimate from the data the true 21-cm optical depth of foreground neutral gas (e.g. \citealt{Braun:2012}). In the subsequent analysis we therefore leave the source covering fraction as a free parameter. 

Previous searches for \mbox{H\,{\sc i}} absorption with ASKAP were towards individual bright radio sources at the phase centre (e.g. \citealt{Allison:2015}). Here we have carried out a search for \mbox{H\,{\sc i}} absorption towards multiple sources across the GAMA\,23 field, and so we must also consider the effects of wide-field imaging. For the 1\,MHz channel-averaged data, the effect of bandwidth smearing at 900\,MHz is comparable to the synthesized beam at angular separations greater than about 5\,deg from the phase centre. Likewise, the ASKAP correlator uses 10\,s time-averaging, which leads to more than 10\,per\,cent reduction in source amplitude at separations greater than about 5\,deg from the phase centre. Compared with the theoretical half power width of the PAF beam, ${\lambda}/D_{\rm dish} \approx 1.6$\,deg, the effects of bandwidth and time-average smearing are therefore negligible. Of greater importance is the effect of non-coplanar antenna baselines; since the INVERT imaging task in \textsc{MIRIAD} does not implement facet gridding (\citealt{Cornwell:1992}) or $w$-term projection (\citealt{Cornwell:2008}) to correct for wide-field aberration, we must account for the degree of distortion of sources away from the phase centre in each PAF beam. For a co-planar array, the apparent position shift in arcseconds, caused by the phase error introduced by ignoring the $w$-term, is approximately equal to $2.4 \times 10^{-6} \theta^{2} \sin{(z)}$, where $\theta$ is the separation from the phase centre in arcseconds and $z$ is the angle from zenith. This error is comparable to the synthesized beam at a separation of about 1\,deg from the phase centre and is evident in the continuum images of sources beyond the half-power point of individual PAF beams. This reduces the point source amplitude, and hence sensitivity to \mbox{H\,{\sc i}} absorption, by approximately 10\,per\,cent across the observed field, but importantly does not introduce any systematic error in our results.

Continuum subtraction was performed on the calibrated, full spectral resolution visibility data in two stages; first we generated a CLEAN component model for each 1-MHz chunk and subtracted it using the MIRIAD task UVMODEL. We then removed residual continuum using the task UVLIN with a second-order polynomial. Since the PAF element weights used to form the beams are applied uniformly in frequency intervals before channelization at full 18.5-kHz resolution, any discontinuous jumps in amplitude and phase are subtracted out through this procedure. However, this necessarily leads to the subtraction of spectral features that have a width greater than the beam forming intervals. Since the early science data presented here were formed in 1-MHz intervals, our results are therefore incomplete to lines broader than approximately 300\,km\,s$^{-1}$. This is large compared with the distribution of line widths for intervening 21-cm absorbers in the literature (see e.g. \citealt{Curran:2016a}), and so unlikely to affect our detecting rate for these systems. However, this could be a factor in detecting intrinsic 21-cm absorbers associated with AGNs, which typically have broader line profiles (see \autoref{section:intrinsic_absorbers}). For the full FLASH survey, we expect to form the PAF beams on larger 10-MHz intervals, equal to line widths more than a few thousand km\,s$^{-1}$. 

The image cubes were generated with natural weighting to optimize sensitivity to absorption line detection, with a subsequent 60\,per\,cent reduction in spatial resolution compared with the full-bandwidth continuum images. We summarize the spectral line flux density sensitivity achieved per 18.5-kHz channel for each observation in \autoref{table:GAMA23_obs}, which is consistent with the duration of observation and number of antennas. Some variation in sensitivity between observations is expected given the process of electronically forming the individual PAF beams. Given that the expected final rms noise in the averaged data is 3\,mJy\,beam$^{-1}$ per 18.5\,kHz, and assuming a 100\,per\,cent source covering fraction, sources with a flux density greater than 10\,mJy are sufficient to detect any absorption at a signal-to-noise ratio (S/N) of better than 3. We used the NVSS catalogue (\citealt{Condon:1998}) as a prior to identify the positions of target continuum sources within the field, which is  complete for source flux densities greater than 2.5\,mJy and has positional accuracy better than 7\,arcsec. Although we could have identified target sources using the ASKAP continuum image, the NVSS is matched in spatial resolution (45\,arcsec) to the data cubes and is therefore suitable to use here. For each PAF beam and observation, the spectra were extracted from the data cube at the position of peak source flux density in the corresponding full-bandwidth ASKAP continuum image. Each spectrum was converted to fractional absorption by dividing by the continuum flux density measured as a function of frequency at that position, using continuum images generated from each 1-MHz band-chunk. We then created a single spectrum for the source by averaging over the spectra using an inverse variance weighting to optimize the S/N. 

In \autoref{figure:GAMA23_mosaic}, we show the full ASKAP-12 image of the 900-MHz continuum in the GAMA\,23 field, which we constructed by carrying out a linear mosaic over all observations and PAF beams. In \autoref{figure:GAMA23_example_spectrum}, we show an example ASKAP-12 spectrum at the position of NVSS\,J224500$-$343030, which demonstrates the spectral fidelity and sensitivity as a function of redshift for these data. The median rms noise in our final spectra is 3.2\,mJy\,beam$^{-1}$ per 18.5-kHz channel, with a standard deviation of 20\,per\,cent across the observed band and field. We note that this is similar to the expected rms noise in a single 2\,h pointing with the FLASH survey (e.g. \citealt{Allison:2016b}) and so these data are a useful early demonstration of the survey. However, the sensitivity to \mbox{H\,{\sc i}} absorption is a function of the source flux density; of our spectra, 1253 have a median rms noise per 18.5\,kHz channel less than 100\,per\,cent fractional absorption of the background continuum, 469 have an rms noise less than 10\,per\,cent and 31 have an rms noise less than 1\,per\,cent - this is summarised for each sightline in \autoref{figure:GAMA23_mosaic}. In the following analysis, we have used all 1253 spectra with a median rms noise less than 100\,per\,cent fractional absorption, so as to detect all possible absorption lines present in the data. The \mbox{H\,{\sc i}} redshift spanned by each spectrum is between $z_{\rm HI} = 0.34$ and $0.79$, but the total absorption search path sensitive to a given minimum column density depends on the redshifts of individual sources, the subtended source flux density, and the \mbox{H\,{\sc i}} spin temperature. We discuss this further in the following analysis.

\section{Results and discussion}\label{section:results}

\begin{figure}
\centering
\includegraphics[width=0.45\textwidth]{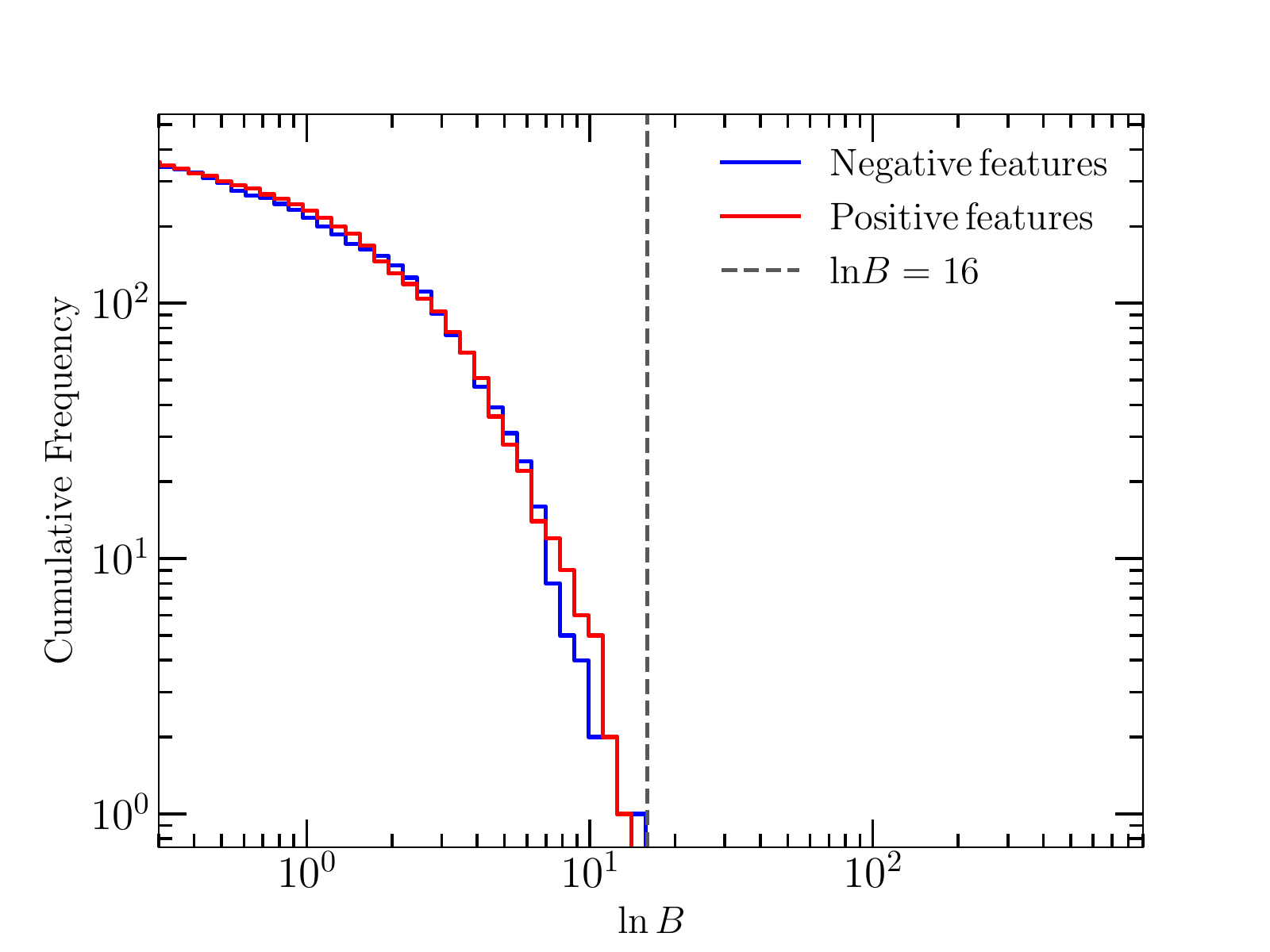}
\caption{The cumulative frequency distribution of spectral features detected in ASKAP-12 observations of the GAMA\,23 field, as a function of the Bayes factor $B$. The blue and red solid lines denote negative and positive spectral features, respectively. Since we do not expect to detect positive features (i.e. emission lines) in our data, we assign a threshold of $\ln{B} > 16$ for reliable detection of absorption (vertical dashed line). As can be seen, above this reliability threshold we do not detect any \mbox{H\,{\sc i}} absorption.}
\label{figure:GAMA23_reliability}
\end{figure} 

\subsection{Detection and reliability}

We search for \mbox{H\,{\sc i}} absorption in our spectra using an automated detection method based on Bayesian model comparison (see \citealt{Allison:2012b, Allison:2014}). This method implements multimodal nested sampling (\citealt{Feroz:2008, Feroz:2009b}) to enable multiple redshifted lines to be found in a given spectrum. The significance of each detected feature is given by the Bayes factor $B$ (e.g. \citealt{Kass:1995}), a statistic that is equal to ratio of Bayesian evidences of a Gaussian absorption-line model and a null model, which contains no line. This assumes no prior preference for either model, which is reasonable if we are testing for the incidence of absorption lines. We note that this method of feature detection is model dependent, but that it is reasonable to expect that extragalactic 21-cm absorption will comprise one or more Gaussian velocity distributions. We use the likelihood function for a normal distribution, with standard deviation equal to the measured rms noise in each channel. The following non-informative priors are used for each model parameter: for the line position we use a uniform prior over the full range of the spectrum, for the FWHM we use a loguniform prior between 0.1 and 2000\,km\,s$^{-1}$, and for the peak optical depth we use a loguniform prior between 1\,per\,cent of the median rms noise and a maximum value of 10. These ranges are chosen to include reasonable boundaries set by the data and physically realistic limits.

Since the data are sparsely populated by an unknown distribution of absorption lines, we expect that the noise will necessarily produce a distribution of spurious detections. We therefore need to establish a reliability threshold for $B$ above which we are confident a given detected feature is real. Given the sensitivity of our data\footnote{The rms noise per channel in our spectra is $(3.2 \pm 0.6)$ \,mJy\,beam$^{-1}$, which for an unresolved galaxy at the lowest redshift  of $z = 0.34$ gives a 5-$\sigma$ detection limit on 21-cm emission that corresponds to an \mbox{H\,{\sc i}} mass of $M_{\rm HI} \sim 4.1 \times 10^{11}$\,M$_{\odot}$ (e.g. \citealt{Meyer:2017}). This is beyond the extreme high-mass end of the local \mbox{H\,{\sc i}} mass function (\citealt{Jones:2018}) and therefore we do not expect any detections of 21-cm emission.} we do not expect to detect any real \mbox{H\,{\sc i}} 21-cm emission lines and can therefore use the distribution of positive features to determine this reliability threshold (see e.g. \citealt{Serra:2012b}). In \autoref{figure:GAMA23_reliability} we plot the cumulative frequency distribution of both negative and positive features as a function of the Bayes factor. A threshold of $\ln(B) > 16$ is found to be sufficient for a reliable detection of absorption in our data\footnote{The reliability of our ASKAP-12 data is limited by a multiplicative non-Gaussian contribution to the noise at the level of 1\,per\,cent. This was due to incorrect firmware weights used in correcting for the coarse channelisation of the data. This error has now been corrected and does not affect data obtained with the full 36-antenna ASKAP.}. Based on this reliability criterion we do not detect any \mbox{H\,{\sc i}} absorption in a blind search of our data. 

\subsection{Sensitivity to HI absorption}

\begin{figure}
\centering
\includegraphics[width=0.45\textwidth]{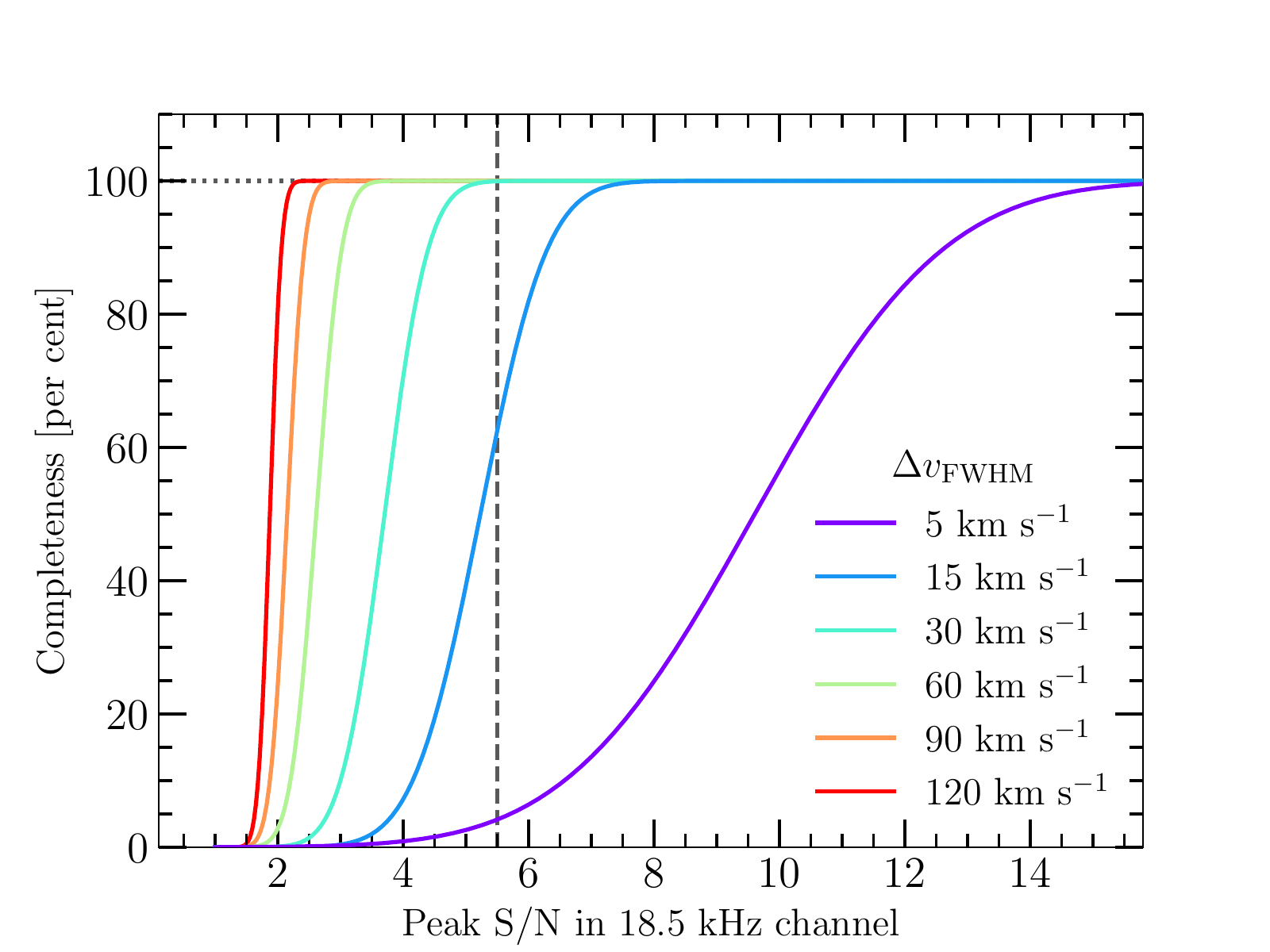}
\caption{The expected completeness of reliable detections in our data as a function of the peak S/N in a single 18.5-kHz channel and the FWHM of the optical depth velocity distribution. The vertical dashed line denotes the peak signal-to-noise ratio at which we recover all lines with FWHM velocity width equal to $\Delta{v}_{\rm FWHM} = 30$\,km\,s$^{-1}$; the mean width for intervening 21-cm absorbers detected in the literature (see e.g. \citealt{Curran:2016a, Allison:2016b}).}
\label{figure:GAMA23_completeness}
\end{figure}

We calculate the 21-cm optical depth sensitivity of each spectral element (i.e. 18.5-kHz channel) in our data using
\begin{equation}\label{equation:total_optical_depth}
    \int\tau_{21}\,{\rm d}v > 0.57 \left[\frac{\tau_{\rm peak}}{0.018}\right]\,\left[\frac{\Delta{v}_{\rm FWHM}}{30\,\mathrm{km}\,\mathrm{s}^{-1}}\right]\mathrm{km}\,\mathrm{s}^{-1},
\end{equation}
where $\Delta{v}_{\rm FWHM}$ is the FWHM rest-frame velocity, assuming a Gaussian distribution, and $\tau_{\rm peak}$ is the peak optical depth sensitivity. For a fixed velocity width, $\tau_{\rm peak}$ is given by
\begin{equation}\label{equation:peak_optical_depth}
\tau_{\rm peak} = -\ln\left(1-0.018\left[\frac{\mathrm{S/N}}{5.5}\right]\,\left[\frac{\sigma_{\rm chan}}{3.2\,\mathrm{mJy\,bm^{-1}}}\right]\,\left[\frac{S_{\rm c}\,c_{\rm f}}{1\,\mathrm{Jy\,bm^{-1}}}\right]^{-1}\right),
\end{equation}
where S/N is the minimum peak signal-to-noise ratio required for recovery of absorption lines to high completeness, $\sigma_{\rm chan}$ is the rms noise, $S_{\rm c}$ is the background continuum flux density and $c_{\rm f}$ is the unknown areal fraction of the unresolved source covered by foreground \mbox{H\,{\sc i}}. In the following, we assume that most 21-cm absorbers are in the optically thin regime ($\tau_{\rm peak} << 1$), so that the peak optical depth is inversely proportional to the source covering fraction and we explicitly give our results as a function of a single unknown; the spin temperature to source covering fraction ratio, $T_{\rm spin}/c_{\rm f}$. However, it should be noted that for the rarest optically thick 21-cm absorbers this approximation does not hold and the source covering fraction should be included in the natural logarithm in \autoref{equation:peak_optical_depth}.

We estimate the required S/N in \autoref{equation:peak_optical_depth} by randomly populating our spectra with 1000 fake absorption lines and calculating the fraction recovered using our detection method and reliability threshold. In \autoref{figure:GAMA23_completeness}, we show the results for velocity widths ranging from $\Delta{v}_{\rm FWHM} = 5$ to 120\,km\,s$^{-1}$. As expected, narrower lines are less complete for a fixed peak sensitivity, but more complete for a fixed total optical depth sensitivity. Since the sensitivity does depend on the choice of velocity width, we adopt a value equal to the mean for all intervening 21-cm absorbers detected in the literature, $\Delta{v}_{\rm FWHM} = 30$\,km\,s$^{-1}$ (see e.g. \citealt{Curran:2016a, Allison:2016b}). In this case, we recover all lines that have a peak S/N greater than 5.5, and that is the value we adopt in the following analysis.

The \mbox{H\,{\sc i}} column density sensitivity of each spectral element is then given by
\begin{equation}\label{equation:column_density_sensitivity}
    N_{\rm HI} > 1.0 \times 10^{20}\,\left[\frac{T_{\rm spin}}{100\,\mathrm{K}}\right]\,\left[\frac{\tau_{\rm peak}}{0.018}\right]\,\left[\frac{\Delta{v}_{\rm FWHM}}{30\,\mathrm{km}\,\mathrm{s}^{-1}}\right]\mathrm{cm}^{-2},
\end{equation}
where the spin temperature ($T_{\rm spin}$) is the column-density-weighted harmonic mean for all phases of the \mbox{H\,{\sc i}} gas along the line of sight through the absorber. In the case of the Galactic ISM, the harmonic mean spin temperature is $T_{\rm spin} \approx 300$\,K with three \mbox{H\,{\sc i}} phases consisting of cold (CNM; $T_{\rm spin} \approx 100$\,K), unstable (UNM; $T_{\rm spin} \approx 500$\,K) and warm neutral medium (WNM; $T_{\rm spin} \approx 10^{4}$\,K), in mass fractions of 28, 20, and 52\,per\,cent, respectively (\citealt{Murray:2018}). In general, the mass fraction of phases is expected to vary depending on the physical conditions within each absorber (see e.g. \citealt{Kanekar:2014a} and references therein). For comparison with previous work in the literature we adopt a fiducial value of $T_{\rm spin} = 100$\,K, but also show results for $T_{\rm spin} = 1000$\,K.  

 We next estimate the total comoving absorption path length ($\Delta{X}$) spanned by our data, by summing over all spectral channels that are sensitive to a minimum column density given by \autoref{equation:column_density_sensitivity}. For the standard $\Lambda$CDM cosmology, the comoving absorption path element for the $i$'th channel is given by 
\begin{equation}
 \Delta{X}_{i} = \Delta{z_{i}}\,(1+z_{i})^{2}\,E(z_{i})^{-1},
\end{equation}
where
\begin{equation}
 E(z_{i}) = \sqrt{(1+z_{i}^{3})\,\Omega_{\rm m} - (1+z_{i})^{2}\,(\Omega_{\rm m} + \Omega_{\Lambda} - 1) + \Omega_{\Lambda}},
\end{equation} 
\begin{equation}
\Delta{z}_{i} = (1+z_{i})\frac{\Delta{\nu}_{\rm chan}}{\nu_{i}},
\end{equation}
\begin{equation}
{z}_{i} = \frac{\nu_{\rm HI}}{\nu_{i}} - 1,
\end{equation}
$\nu_{i}$ is the solar barycentric-corrected centre frequency of the $i$'th channel, $\Delta{\nu}_{\rm chan}$ is the channel separation, equal to 18.5\,kHz, and $\nu_{\rm HI}$ is the rest frequency of the \mbox{H\,{\sc i}} 21-cm line, equal to 1420.40575177\,MHz (\citealt{Hellwig:1970}).

\begin{figure}
\centering
\includegraphics[width=0.475\textwidth]{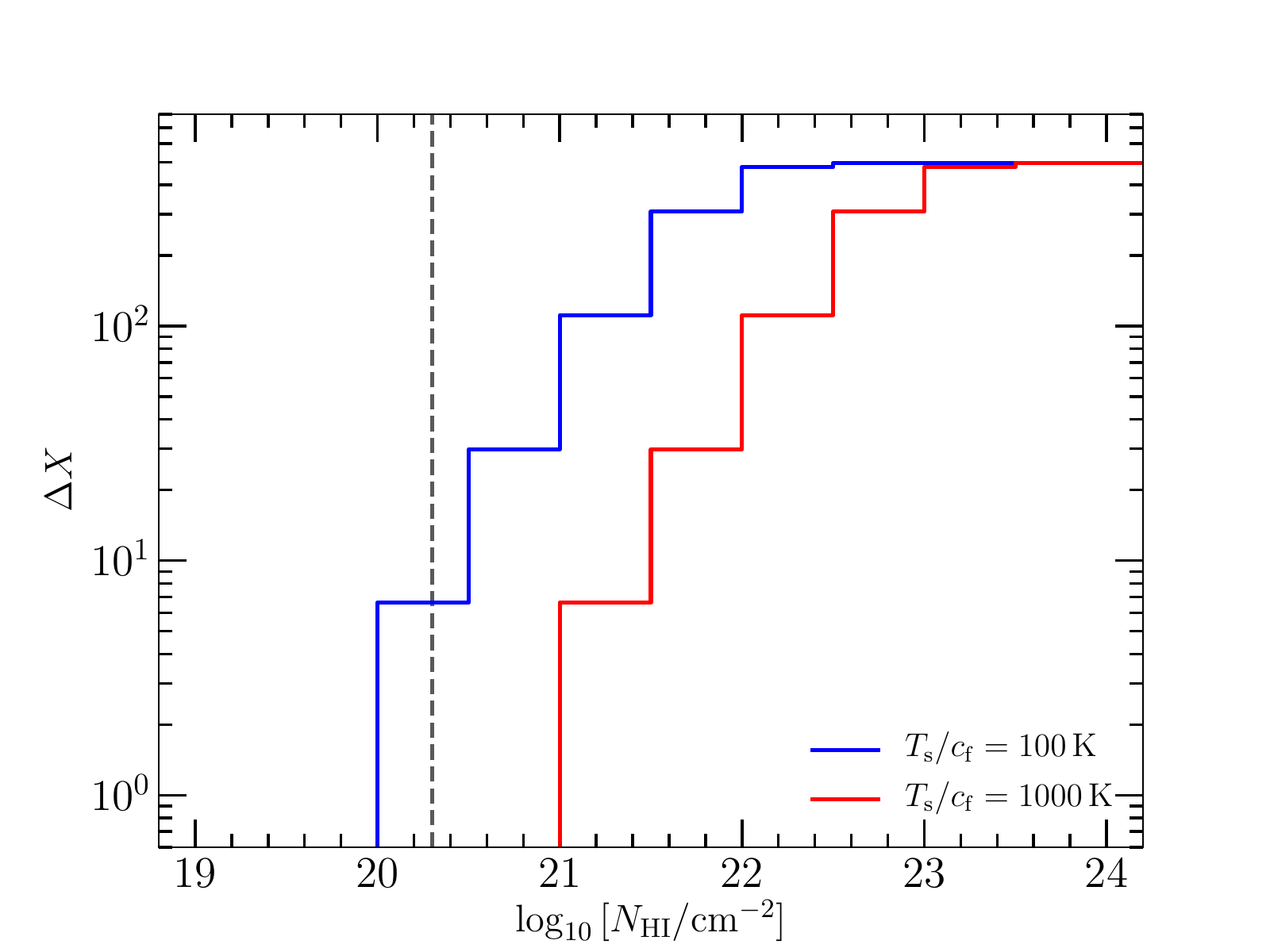}
\caption{The comoving absorption path length ($\Delta{X}$) spanned by our data as a function of \mbox{H\,{\sc i}} column density sensitivity. We give results for spin temperature to source covering fraction ratios of 100 (blue line) and 1000\,K (red line). The vertical dashed line indicates the lower limit definition for DLAs ($N_{\rm HI} \geq 2 \times 10^{20}$\,cm$^{-2}$).}
\label{figure:GAMA23_nhi_sensitivity}
\end{figure}

The comoving absorption path length covered by each source spectrum depends on the cosmological redshift of that source, which in some cases may be within the range spanned by our data or even at a lower redshift. We account for this by weighting each $\Delta{X}_{i}$ by the probability that the source lies at a cosmological redshift greater than the spectral channel. We exclude absorption associated with \mbox{H\,{\sc i}} gas in the host galaxy of the radio source by shifting the channel redshift used in the weighting by a rest-frame radial velocity of $\Delta{v}_{\rm asc} = 3000$\,km\,s$^{-1}$. The total comoving absorption path length is thus given by the following sum over all spectral channels that are sensitive to a given minimum column density
\begin{equation}
\Delta{X} = \sum_{i}w_{i}\,\Delta{X}_{i},	
\end{equation}
where 
\begin{equation}\label{equation:redshift_weighting}
	w_{i} = w(z_{i}+ \Delta{z}_{\mathrm{asc}, i}),
\end{equation}
\begin{equation}
	\Delta{z}_{\mathrm{asc}, i}= (1 + z_{i})\,\Delta{v}_{\rm asc}/c,
\end{equation}
and $w$ is the redshift weighting function.

We do not \emph{a\,priori} use optical matching to confirm redshifts for individual sources, which would be both highly incomplete for our large sample and prone to confusion error. Instead we use a statistical weighting for all sources based on the redshift distribution given by \cite{deZotti:2010} from a fit to bright ($S_{\rm 1.4} \geq 10$\,mJy) sources in the Combined EIS-NVSS Survey Of Radio Sources (CENSORS; \citealt{Brookes:2008}). In this case, the weighting function used in \autoref{equation:redshift_weighting} is given by 
\begin{equation}
w(z) = \frac{\int_{z}^{\infty}{\mathcal{N}_{\rm src}(z^{\prime})\,\mathrm{d}z^{\prime}}}{\int_{0}^{\infty}{\mathcal{N}_{\rm src}(z^{\prime})\,\mathrm{d}z^{\prime}}},
\end{equation}
where the redshift distribution is given by (\citealt{deZotti:2010})
\begin{equation}
\mathcal{N}_{\rm src}(z) = 1.29 + 32.37\,z - 32.89\,z^{2} \
 + 11.13\,z^{3} - 1.25\,z^{4}.	
\end{equation}
For the \mbox{H\,{\sc i}} redshifts covered by our data, the fractional uncertainty in the cumulative redshift distribution due to the size of the CENSORS sample is approximately $3 - 6$\,per\,cent. This translates directly to the same fractional uncertainty in our estimates of the total comoving absorption path length and expected detection rates for the survey. 

In \autoref{figure:GAMA23_nhi_sensitivity} we show the total comoving absorption path length as a function of \mbox{H\,{\sc i}} column density. We show results for $T_{\rm spin}/c_{\rm f} = 100$ and $1000$\,K, which are indicative values spanning that typically measured for sight lines through the Milky Way interstellar medium (e.g. \citealt{Heiles:2003b, Murray:2018}). The total comoving absorption path length spanned by our data is $\Delta{X} = 500 \pm 25$, and for $T_{\rm spin}/c_{\rm f} = 100$\,K the total path sensitive to all DLAs is $\Delta{X} = 6.6 \pm 0.3$ (redshift interval of $\Delta{z} = 3.7 \pm 0.2$) and super-DLAs ($N_{\rm HI} \geq 2 \times 10^{21}$\,cm$^{-2}$) is $\Delta{X} = 111 \pm 6$ ($\Delta{z} = 63 \pm 3$).

\begin{figure}
\centering
\includegraphics[width=0.5\textwidth]{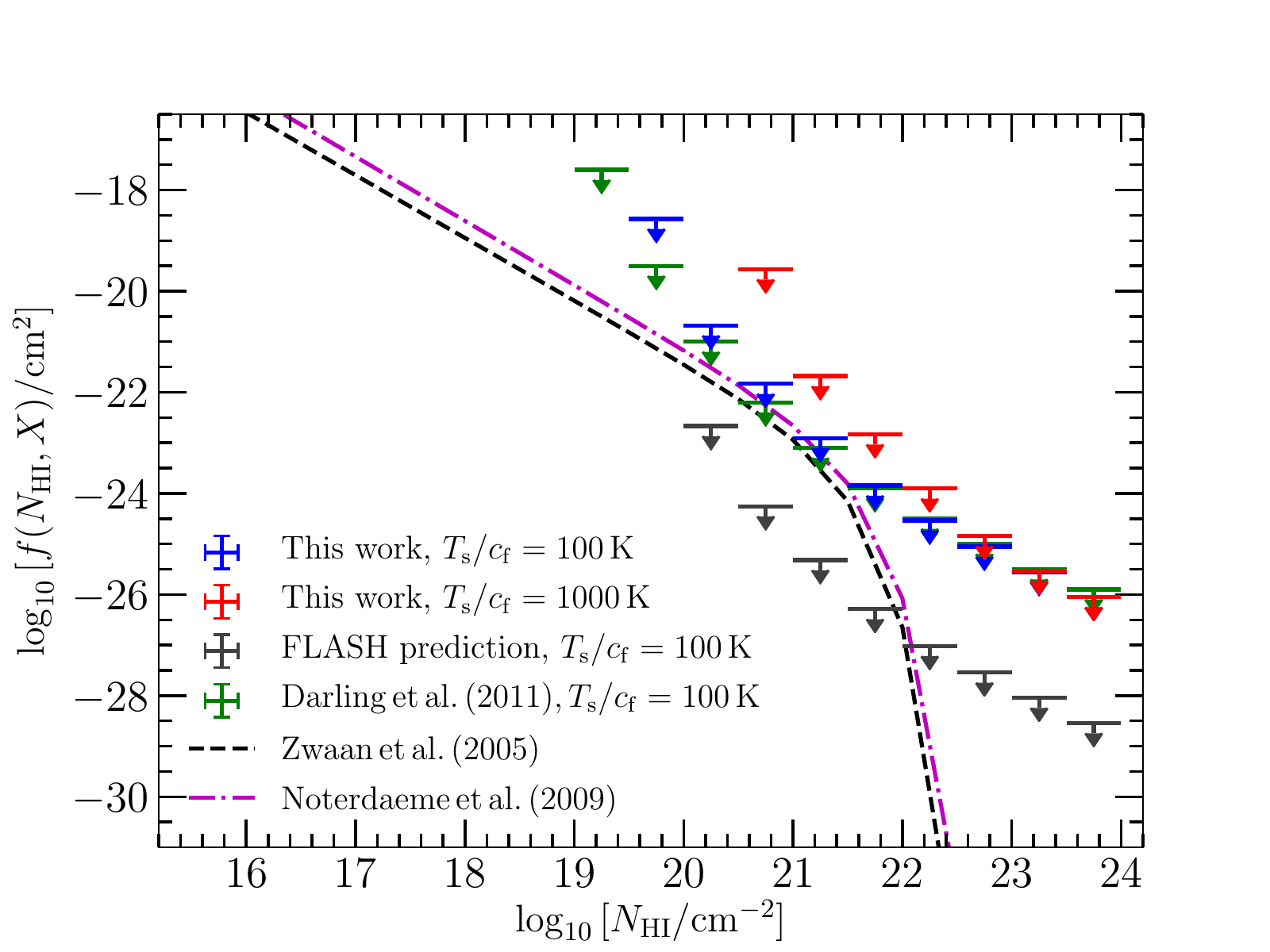}
\caption{95\,per\,cent upper limits on the \mbox{H\,{\sc i}} column density frequency distribution from our ASKAP-12 observations of the GAMA23 field. We give distributions for values of the spin temperature to source covering fraction ratio of 100\,K (blue points) and 1000\,K (red points). We predict the sensitivity of the planned FLASH survey (assuming $T_{\rm spin}/c_{\rm f} = 100$\,K) by scaling the integration time (2\,h), number of antennas (36), sky area (south of $\delta = +10$\,deg) and redshift coverage ($z = 0.4$ to $1.0$). Also shown are results from the ALFALFA absorption pilot survey by \citet{Darling:2011} and model fits to data given by \citet{Zwaan:2005} at $z = 0$ and \citet{Noterdaeme:2009} at $z \approx 3$.}
\label{figure:GAMA23_nhi_distribution}
\end{figure}

% \subsection{Comparison with previous 21-cm absorption surveys}

% Although we have not detected any 21-cm absorbers in a blind search of the GAMA\,23 field, we can determine if this is consistent with previous estimates of the incidence rate. There have been several prior surveys for intervening 21-cm absorbers that have targeted known DLA and MgII absorbers in the visible and ultraviolet bands. Here we consider the most recent and complete samples to ensure a useful comparison. In a study of 37 DLAs with measurements of 21-cm absorption, \cite{Kanekar:2014a} obtained an incidence rate of $83^{+17}_{-21}$\,per cent for systems at redshifts below the sample median $z_{\rm DLA} = 2.192$. 

\subsection{Limits on the $N_{\rm HI}$ frequency distribution function and spin temperature}

As can be seen in \autoref{figure:GAMA23_nhi_sensitivity}, our data vary by several orders of magnitude in \mbox{H\,{\sc i}} column density sensitivity and as a function of the spin temperature and source covering fraction. To compare our results with previous 21-cm and DLA surveys for \mbox{H\,{\sc i}} gas, it is therefore necessary that we determine the sensitivity of our data to the frequency of intervening systems as a function of \mbox{H\,{\sc i}} column density and $T_{\rm spin}/c_{\rm f}$. Upper limits on the $N_{\rm HI}$ frequency distribution function, $f(N_{\rm HI}, X)$, are given by
	\begin{equation}
	 	f(N_{\rm HI}, X) < \frac{\lambda_{\rm max}}{\Delta{N_{\rm HI}}\Delta{X}},
	 \end{equation}
where $\lambda_{\rm max}$ is the Poisson upper limit on the detection rate of absorbers with column density $N_{\rm HI}$ in interval $\Delta{N_{\rm HI}}$ and $\Delta{\rm X}$ is the total comoving absorption path length sensitive to $N_{\rm HI}$. We use $\Delta{N_{\rm HI}} = 0.5$\,dex and the 95\,per\,cent upper limit on the Poisson rate given by $\lambda_{\rm max} = 3$ when no detections are obtained. This allows us to compare our limits directly with that of \cite{Darling:2011}, who carried out a blind 21-cm absorption survey of the nearby Universe using pilot data from the Arecibo Legacy Fast Arecibo $L$-band Feed Array (ALFALFA) survey (\citealt{Giovanelli:2005}). 

We show our upper limits on $f(N_{\rm HI}, X)$ in \autoref{figure:GAMA23_nhi_distribution}, for spin temperature to source covering fraction ratios of $T_{\rm spin}/c_{\rm f} = 100$ and 1000\,K. For super-DLA column densities ($N_{\rm HI} \geq 2 \times 10^{21}$\,cm$^{-2}$) these are similar to that of \cite{Darling:2011}, but less sensitive at lower column densities due to the relative flux density sensitivity of the two surveys. We also show the expected sensitivity for the all-sky FLASH survey, scaling to the expected integration time per pointing (2\,h), the number of antennas (36), the sky area (south of $\delta = +10\,\deg$) and the redshift coverage ($z_{\rm HI} = 0.4$ to 1.0). At DLA column densities, FLASH is expected to be two orders of magnitude more sensitive to $f(N_{\rm HI}, X)$ than these early results. 

\begin{figure}
\centering
\includegraphics[width=0.5\textwidth]{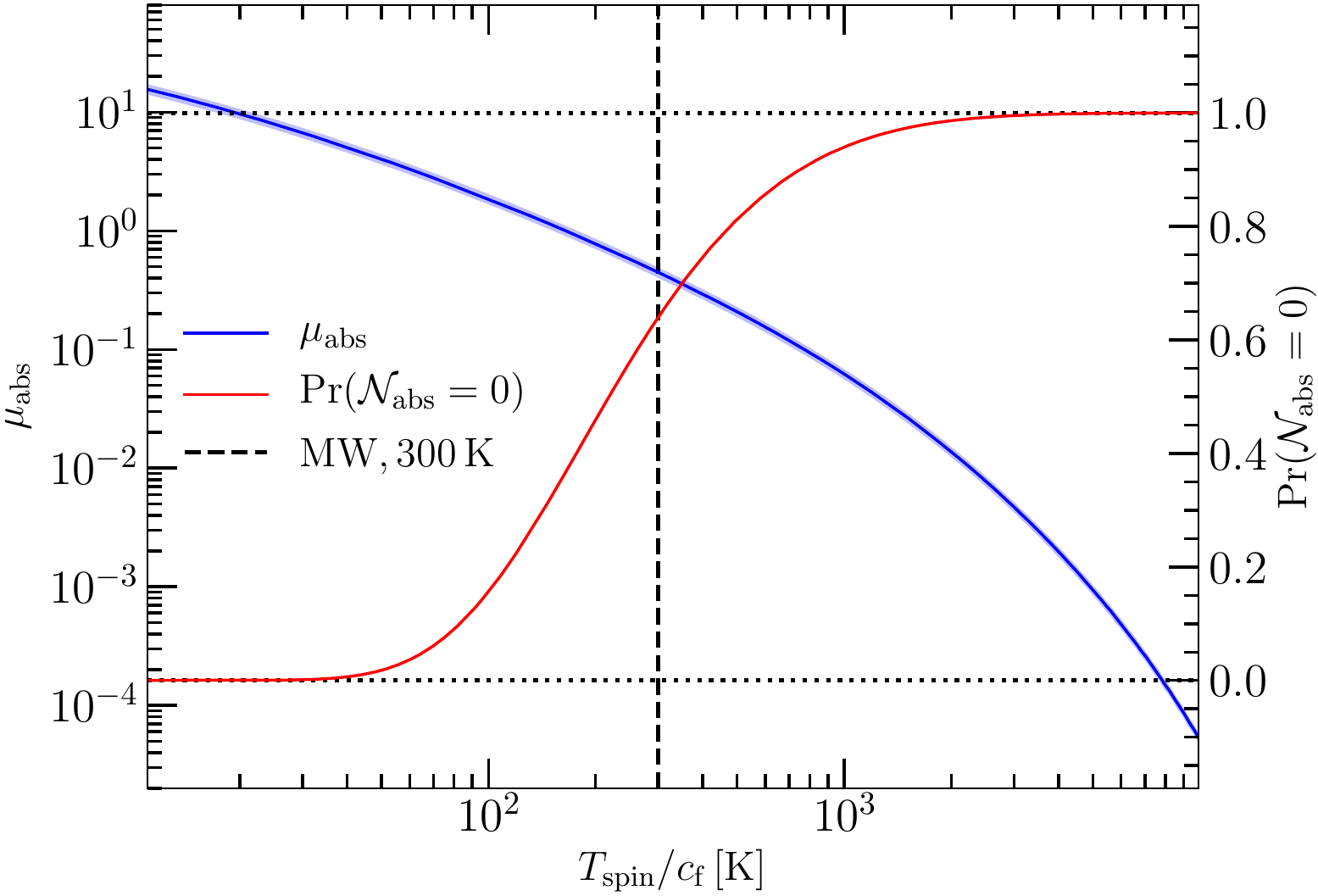}
\caption{The expected number of reliable detections of 21\,cm absorbers in our data ($\mu_{\rm abs}$), and the corresponding Poisson probability of detecting no absorbers, as a function of the spin temperature to source covering fraction ratio. For comparison, also shown is the harmonic mean spin temperature of the Milky Way ISM, assuming $c_{\rm f} = 1$ (\citealt{Murray:2018}).}
\label{figure:GAMA23_ndetections}
\end{figure}

For $T_{\rm spin}/c_{\rm f} \gtrsim 100$\,K, our upper limits on $f(N_{\rm HI}, X)$ are consistent with the measured \mbox{H\,{\sc i}} column density frequency distributions from surveys of 21-cm emission in nearby galaxies (\citealt{Zwaan:2005}) and distant ($z \approx 3$) DLAs (\citealt{Noterdaeme:2009}). Using these measured $f(N_{\rm HI}, X)$, we can calculate the expected number of 21-cm absorber detections in our data, given by
\begin{equation}
	\mu_{\rm abs} = \sum_{i}\,\left[\int_{N_{\rm min}}^{\infty} f(N_{\rm HI}, X_{i})\,\mathrm{d}N_{\rm HI}\right]\,w_{i}\,\Delta{X}_{i},
\end{equation}
where $N_{\rm min}$ is the minimum column density sensitivity given by \autoref{equation:column_density_sensitivity}. We interpolate $f(N_{\rm HI}, X)$  over the redshift range covered by our observations; since the total \mbox{H\,{\sc i}} content of the Universe is not expected to evolve strongly at these redshifts, our results are not sensitive to the degree of interpolation employed. The typical measurement uncertainty in $f(N_{\rm HI}, X)$, due to the finite sample sizes of galaixes, is 10\,per\,cent over the range of column densities to which our ASKAP data are sensitive. Combining this with the uncertainty in $w(z)$ due to the source redshift distribution, we estimate a fractional uncertainty of 11\,per\,cent in $\mu_{\rm abs}$.

In \autoref{figure:GAMA23_ndetections}, we show the expected number of 21-cm absorber detections as a function of $T_{\rm spin}/c_{\rm f}$, and the corresponding Poisson probability of obtaining zero detections. For the harmonic mean spin temperature of the Milky Way ISM, $T_{\rm spin} \approx 300$\,K (assuming $c_{\rm f} = 1$), the probability of detecting zero absorbers is 64\,per\,cent. Our data are consistent with a mean spin temperature of \mbox{H\,{\sc i}} gas at intermediate cosmological distances ($z \sim 0.5$) equal to the Milky Way ISM, but with a significant non-zero probability for colder gas; the likelihood of $T_{\rm spin}/c_{\rm f} \lesssim 100$\,K for our data is 16\,per\,cent. Any evolution in the CNM fraction at these redshifts will be confirmed by future all-sky surveys that should provide an order of magnitude stronger constraint on the spin temperature as a function of redshift (see e.g. \citealt{Allison:2016b}).

\subsection{Intrinsic 21-cm absorbers}\label{section:intrinsic_absorbers}

In addition to intervening galaxies, 21-cm absorption is also detectable from \mbox{H\,{\sc i}} gas in the host galaxy of the radio source  (see \citealt{Morganti:2018} for a review). Since we did not detect any 21-cm absorbers within the volume of our survey, it is instructive to compare this result with the incidence of detections reported in recent large-scale targeted surveys for intrinsic 21-cm absorbers. The expected detection rate of intrinsic absorbers in active galaxies is not well defined because of the complex relationship between the physical conditions of the neutral gas in the host galaxy and the AGN. In particular, there is a factor of $\sim 6$ discrepancy between the detection rates in nearby objects and those at higher redshifts (e.g. \citealt{Curran:2008, Curran:2010, Curran:2012a, Aditya:2016, Aditya:2018b,  Curran:2019b, Grasha:2019}). The lower detection rates of intrinsic absorbers in high-redshift objects are thought to be due to selection bias -- from either UV-luminous quasars that ionize the gas or excitation of the gas by 21-cm photons from the radio source  -- and possibly also intrinsic redshift evolution in the host galaxy population. In the future, planned wide-field radio-selected surveys using the SKA and its precursors will be key in studying the \mbox{H\,{\sc i}} content of these distant radio-AGN hosts (\citealt{Morganti:2015}).  

Here we compare our results with that of the radio-selected survey by \cite{Maccagni:2017}. Using the Westerbork Synthesis Radio Telescope (WSRT), they carried out a search for intrinsic \mbox{H\,{\sc i}} absorption in 248 nearby ($z < 0.25$) radio galaxies with core flux densities  $S_{1.4} > 30$\,mJy, reporting a detection rate of ($27 \pm 5.5$)\,per\,cent for their whole sample, with an average peak optical depth sensitivity of approximately $\tau_{\rm peak} \approx 0.05$. The FWHM of line profiles in intrinsic 21-cm absorbers are typically broader, with an average value of $\Delta{v}_{\rm FHWM} \approx 120$\,km\,s$^{-1}$ (see e.g. \citealt{Gereb:2015, Curran:2016a, Maccagni:2017}). For these broader line profiles our detection threshold reliably recovers all absorption lines with a peak S/N greater than 3 in a single 18.5\,kHz channel (see \autoref{figure:GAMA23_completeness}). Of the 1253 radio sources in our sample, 54 sight lines have a median rms optical depth noise per spectral channel less than the 1.67\,per\,cent required to detect $\tau_{\rm peak} = 0.05$ at S/N = 3. 

Given that the sources in our sample are at higher redshifts than that of \cite{Maccagni:2017}, for the same optical depth sensitivity we select for higher radio luminosity. It is possible that this might select against detection of absorption if the \mbox{H\,{\sc i}} gas is excited to higher spin temperatures by 21-cm photons from the radio source (\citealt{Bahcall:1969}). Of the 54 sightlines in our data that have optical depth sensitivity comparable to that of Maccagni et al. the mean and maximum 1.4-GHz flux densities are $\langle{S_{1.4}\rangle} = 340$\,mJy and $\mathrm{max}(S_{1.4}) = 1.3$\,Jy. At the highest redshift covered by our survey, $z = 0.79$, these flux densities correspond to radio luminosities equal to $L_{1.4} = 5.5 \times 10^{26}$ and $2.1 \times 10^{27}$ W\,Hz$^{-1}$, respectively. The sample of \cite{Maccagni:2017} spanned comparatively lower luminosities between $L_{1.4} = 3.2 \times 10^{22}$ and $1.6 \times 10^{26}$\,W\,Hz$^{-1}$, indicating that there could be a bias. However, \cite{Aditya:2018b} obtained a similar detection rate of $28^{+10}_{-8}$\,per\,cent to Maccagni et al. for a sample of 46 intermediate-redshift ($z < 1.2$) compact flat-spectrum radio sources with maximum radio luminosity $\mathrm{max}(L_{1.4}) = 5.0 \times 10^{27}$\,W\,Hz$^{-1}$. Furthermore, in their targeted survey of 145 compact radio sources, spanning luminosities in the range $L_{1.4} = 1.6 \times 10^{25}$ and $4.5 \times 10^{28}$\,W\,Hz$^{-1}$, \cite{Grasha:2019} found no evidence for a statistical correlation between their detection rate and radio luminosity. Similarly, \cite{Curran:2019b} carried out a study of intrinsic absorbers reported in the literature and found no conclusive evidence that the radio luminosity has a strong effect on the detection rate. Therefore, our results are unlikely to be biased against detecting 21-cm absorbers as a result of our selection for optical depth sensitivity. 

Using the global redshift distribution of sources brighter than 10\,mJy given by \cite{deZotti:2010}, we expect approximately 25\,per\,cent of our sources to be within the volume bounded by $z = 0.34$ to $0.79$. This means that of the 54 sufficiently sensitive sight lines, approximately 14 are expected to be suitable to search for intrinsic 21-cm absorption. As discussed in \autoref{section:observations_data}, our continuum subtraction process removes spectral artefacts on scales greater than the beam forming interval, so that our search is highly incomplete to lines wider than approximately 300\,km\,s$^{-1}$. \cite{Maccagni:2017} find that approximately 25\,per\,cent of their absorption lines have a full-width at 20\,per\,cent maximum (FW20) greater than this velocity width. Assuming a detection rate equal to 27\,per\,cent, and taking into account an additional missing 25\,per\,cent of lines, the binomial probability of detecting exactly zero absorbers in our data is  $\mathrm{Pr}(\mathcal{N}_{\rm abs} = 0) \approx 0.04$. 

There is therefore tension between non-detection of intrinsic absorbers in our data and the detection rate for the whole sample of \cite{Maccagni:2017}.  An important factor in determining the detection rate of intrinsic absorbers is the morphology of the radio source; simply because compact radio sources are more likely to be located within their host galaxy and subtended by higher column densities of absorbing gas (e.g. \citealt{Pihlstrom:2003,Curran:2013b}). Indeed, \cite{Maccagni:2017} report detection rates of $(32 \pm 7.9)$\,per\,cent for the 131 compact sources versus $(16 \pm 6.8)$\,per\,cent for the 108 extended sources in their sample, using a morphological classification based on the major-to-minor axis ratio in NVSS  (\citealt{Condon:1998}) and the peak-to-integrated flux density ratio in the Faint Images of the Radio Sky at Twenty Centimeters (FIRST; \citealt{Becker:1995}). The lower detection rate reported for extended radio sources is more consistent with our data, with a probability of non-detection $\mathrm{Pr}(\mathcal{N}_{
\rm abs} = 0) \approx 0.17$. Since Maccagni at al. selected sources based on the radio core flux density, primarily so as to study the kinematics of \mbox{H\,{\sc i}} absorption, this will bias their whole sample towards higher detection rates. Our results show that a lower rate of detection for intrinsic 21-cm absorbers is expected from future wide-field radio-selected surveys, although by how much is yet to be determined. 

\subsection{Cross-matching with optically identified galaxies}\label{section:gamma_cross_match}

Beyond a simple spectroscopically blind survey of the GAMA\,23 field, we cross-matched all 1253 radio sources in our full sample against the existing spectroscopic catalogue of galaxies in this field (Driver et al. in preparation). For each radio source, we determine a matched galaxy if: (a) it has a spectroscopic redshift within the range for \mbox{H\,{\sc i}} covered by the ASKAP-12 spectrum, and (b) within an impact parameter of 50\,kpc, which is the maximum galacto-centric radius typically seen for \mbox{H\,{\sc i}} discs (e.g. \citealt{Wang:2016}, see also \citealt{Bland-Hawthorn:2017}) and within which 21-cm absorption is expected to be detected (e.g. \citealt{Reeves:2016, Curran:2016c, Borthakur:2016, Dutta:2017}). Note that we do not use available optical photometry to predetermine the presence and/or size of any \mbox{H\,{\sc i}} disc, opting instead for a minimally informative prior based on just the presence of an optical galaxy with a reliable redshift. Based on these criteria, we obtain 51 successful matches between radio sources and GAMA galaxies. The impact parameters range from 0.8 to 50\,kpc over a roughly uniform distribution, with 15 matches within 10\,kpc. We then re-run our line finding method using a normal prior for the line position given by the GAMA redshift. 

\begin{figure}
\centering
\includegraphics[width=0.45\textwidth]{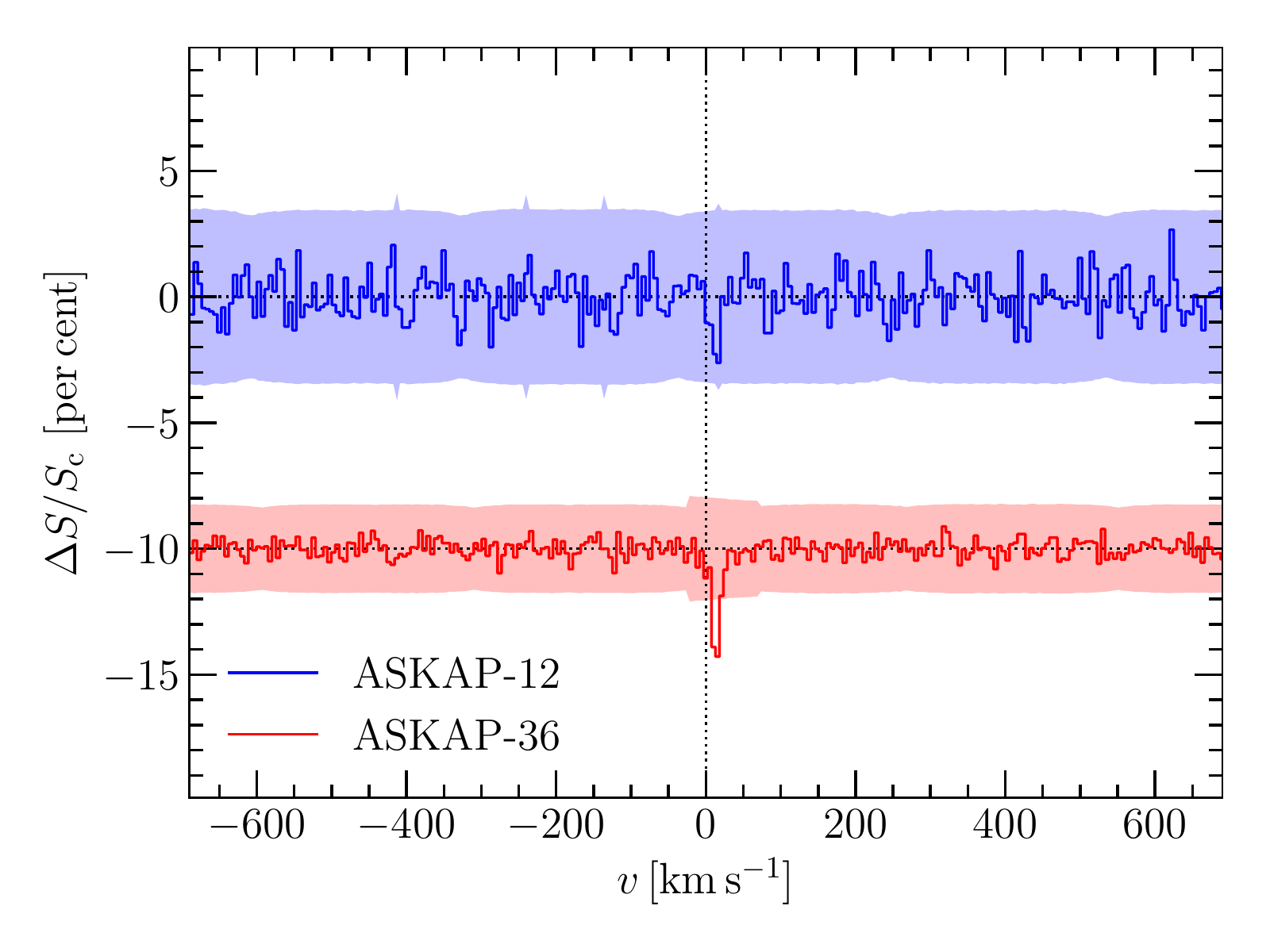}
\caption{ASKAP spectra towards the radio source NVSS\,J224500$-$343030 at the spectroscopically determined redshift ($z = 0.3562$) of the galaxy GAMA\,J224500.05$-$343031.7. The detection of $\mbox{H\,{\sc i}}$ absorption evident in the ASKAP-12 spectrum was later confirmed using the ASKAP-36 array. The continuum-subtracted data (solid lines) are given as a fraction of the  continuum flux density. The shaded regions denote 5 times the rms noise measured from the channel images. For clarity the ASKAP\,36 spectrum has been offset by 10\,per\,cent.}
\label{figure:J224500-343030_xmatch_spectrum}
\end{figure}

Of these 51 cross-matches, we obtain one tentative detection of \mbox{H\,{\sc i}} absorption towards the radio source NVSS\,J224500$-$343030 ($S_{843} = 587.3 \pm 17.7$\,mJy; \citealt{Mauch:2003}) at the redshift of the galaxy GAMA\,J224500.05$-$343031.7 ($z = 0.356$; Driver et al. in preparation). The Bayes factor for the single-component Gaussian model versus noise-only model was $\ln(B) = 3.9 \pm 0.1$, with a corresponding two-model probability for the spectral line of $\mathrm{Pr}(M|D) = 98.0(1)$\,per\,cent. 
This detection was then confirmed by carrying out a further 6\,h observation on 2019 May 17 (SBID 8808) with the full 36-antenna array (ASKAP-36), achieving an rms noise of 2.1\,mJy\,beam$^{-1}$ per 18.5\,KHz channel with a continuum spatial resolution at the centre of the band of $22 \times 16$\,arcsec. This observation used just a single PAF beam centred on the target source. We show both the ASKAP-12 and ASKAP-36 spectra for NVSS\,J224500$-$343030 in \autoref{figure:J224500-343030_xmatch_spectrum}, centred at the redshift of the galaxy. We obtain $\ln(B) = 107.2 \pm 0.1$ for the single-component Gaussian model and find that further complexity with a two-component model is disfavoured with $\Delta{\ln(B)} = -1.8 \pm 0.1$. As an average across the unresolved continuum flux density, we estimate a peak optical depth of $\tau_{\rm peak} = 0.043 \pm 0.004$, width of $\Delta{v}_{\rm FHWM} = (14.1 \pm 1.6)$\,km\,s$^{-1}$ and an integrated optical depth of $\int{\tau_{21}\,\mathrm{d}v} = (0.64 \pm 0.06)$\,km\,s$^{-1}$. The corresponding \mbox{H\,{\sc i}} column density is $N_{\rm HI} = (1.2 \pm 0.1) \times 10^{20}\,(T_{\rm spin}/100\,\mathrm{K})$\,cm$^{-2}$, demonstrating that inclusion of a prior optical identification has enabled us to detect absorption below the sensitivity limit of our blind survey (as shown in \autoref{figure:GAMA23_nhi_sensitivity}).

\subsection{\mbox{H\,{\sc i}} absorption in GAMA J224500.05-343031.7}

\begin{figure}
\centering
\includegraphics[width=0.475\textwidth]{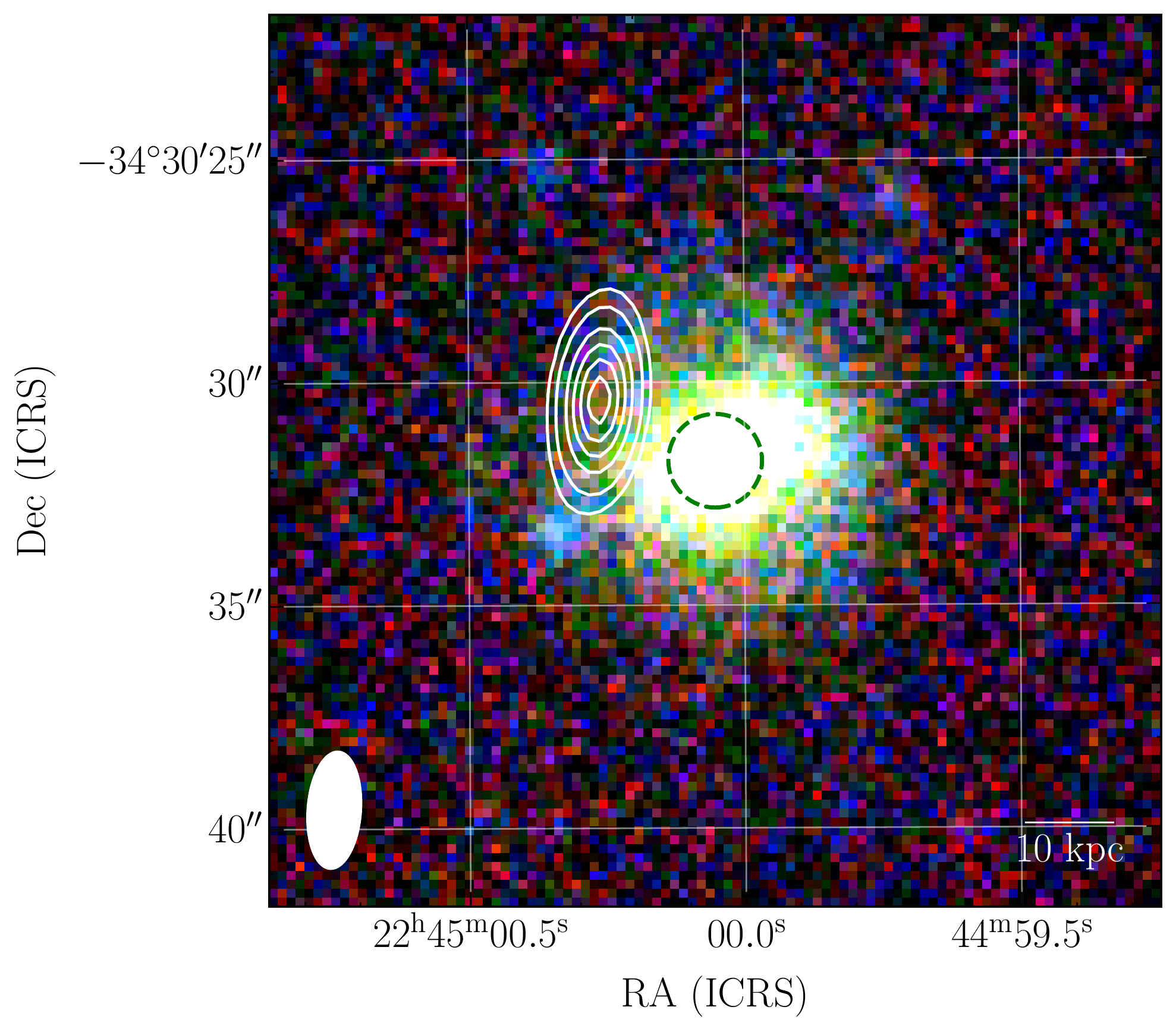}
\caption{A three-colour optical image of GAMA\,J224500.05$-$343031.7, constructed using $gri$-band images from Data Release 4 of the VST Kilo-Degree Survey (KiDS; \citealt{Kuijken:2019}). The dashed green circle indicates the position and width of the GAMA spectroscopic aperture. The white contours show the 9.5\,GHz continuum from ATCA observations of NVSS\,J224500$-$343030 (levels are 5, 10, 20, 30, 40 and 50\,mJy\,beam$^{-1}$). The ATCA restoring beam is shown in the bottom left hand corner. The horizontal bar indicates the physical scale at the redshift of the galaxy.}
\label{figure:J224500-343030_rgb_image}
\end{figure}

\begin{figure}
\centering
\includegraphics[width=0.45\textwidth]{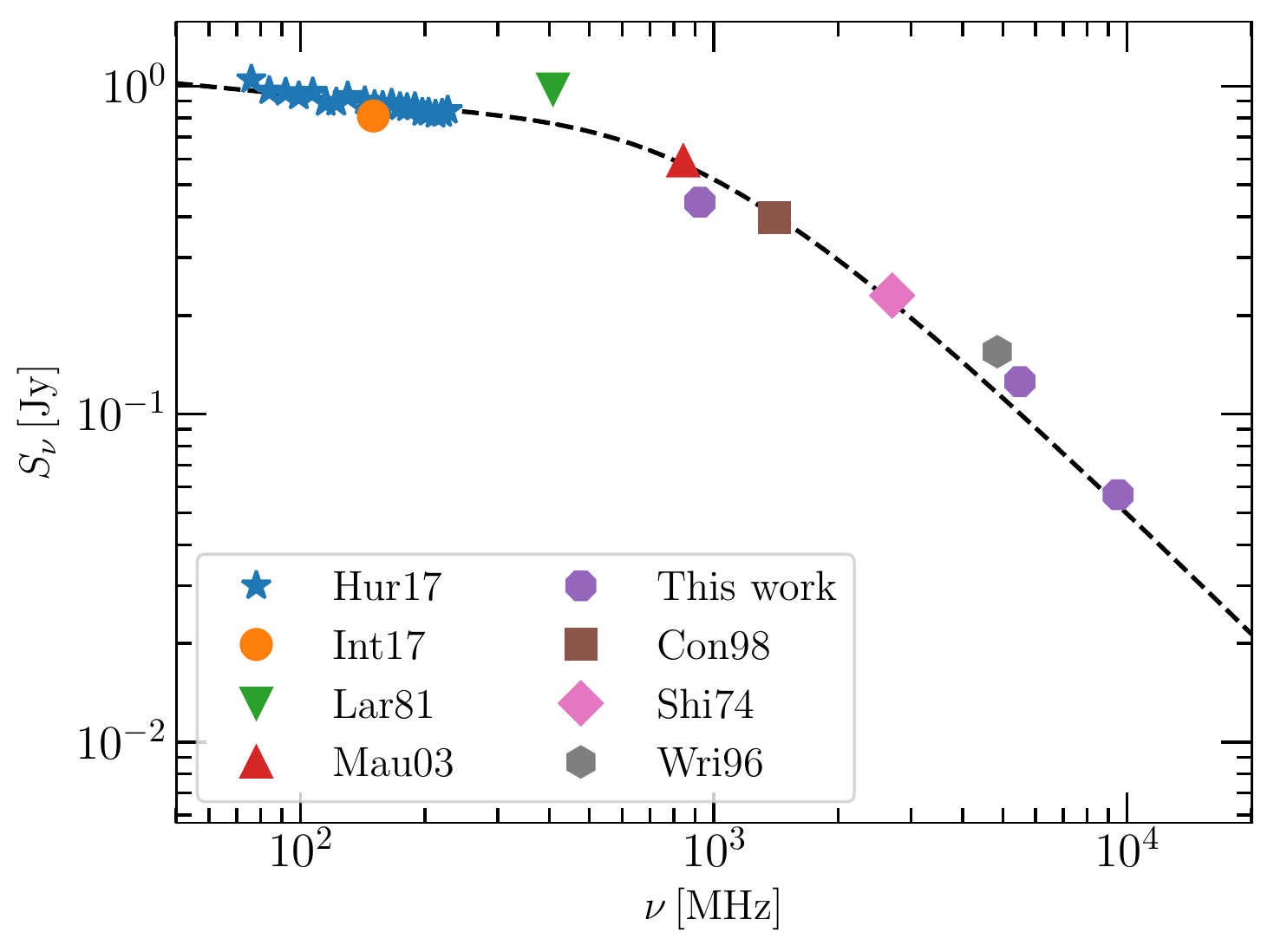}
\caption{The spectral energy distribution (SED) at radio wavelengths of the source NVSS\,J224500$-$343030, compiled using data from the literature data this work. The frequency axis is given in the observer rest frame. The dashed line denotes a best-fitting model that includes optically thick and thin power-law spectra. References for the data: Hurl17 -- \citet{Hurley-Walker:2017}; Int17 -- \citet{Intema:2017}; Lar81 -- \citet{Large:1981}; Mau03 -- \citet{Mauch:2003}; Con98 -- \citet{Condon:1998}; Shi74 -- \citet{Shimmins:1974}; Wri96 -- \citet{Wright:1996}.}
\label{figure:J224500-343030_radio_SED}
\end{figure}

Using ASKAP we have detected \mbox{H\,{\sc i}} absorption at $z = 0.3562$ in the galaxy GAMA\,J224500.05$-$343031.7. In \autoref{figure:J224500-343030_rgb_image} we show a three-colour optical image of this galaxy using $gri$-band images from Data Release 4 of the VST Kilo-Degree Survey (KiDS; \citealt{Kuijken:2019}). The catalogued position of the radio source NVSS\,J224500$-$343030 is offset from the centre of the galaxy by 2.6\,arcsec, which is significant when compared with the uncertainty due to the noise ($\approx 26$\,mas) and calibration in NVSS ($\approx 0.5$\,arcsec; \citealt{Condon:1998}), and the known astrometric error of the VLA with respect to optical surveys ($\approx 20$\,mas; e.g. \citealt{Helfand:2015}). At the redshift of the galaxy this offset is equivalent to a physical separation of 13\,kpc and beyond the half-light radius of $10$\,kpc (1.9\,arcsec), suggesting that the radio source is not associated with the galaxy nucleus. This interpretation is supported by the spectral behaviour of the source, which is characteristic of a compact radio galaxy (see \autoref{figure:J224500-343030_radio_SED}). It has a steep power-law spectrum at frequencies above 1.3\,GHz, with a spectral index of $\alpha \approx -1.2$, that significantly flattens to $\alpha \approx -0.1$ at lower frequencies, likely due to synchrotron self-absorption associated with a characteristic scale of $d_{\rm src} \lesssim 10$\,kpc (e.g. \citealt{Snellen:2000a}). This suggests that the \mbox{H\,{\sc i}} absorption in GAMA\,J224500.05$-$343031.7 is seen towards a background radio source (with a radio luminosity $L_{\rm 5\,GHz} \gtrsim 2 \times 10^{42}$\,erg\,s$^{-1}$). 

In order to confirm that the source is a background AGN, we use higher resolution data of the radio continuum at 5.5 and 9.5\,GHz from the GAMA Legacy ATCA Southern Survey (GLASS; Huynh et al. in preparation)\footnote{\url{https://research.csiro.au/glass/}}. Observations were carried out using the Australia Telescope Compact Array  (ATCA) in $5 \times 40$\,s scans on 2019 May 13 under OPAL\footnote{\url{https://opal.atnf.csiro.au}} project code C3132. The array was in 1.5\,B configuration, which for a Brigg's robustness weighting parameter of 0.5 gave an effective resolution of $4.8 \times 2.0$ and $2.6 \times 1.2$\,arcsec at 5.5 and 9.5\,GHz, respectively. The visibility data were flagged, calibrated, and imaged using standard tasks from \textsc{MIRIAD}. The 5.5 and 9.5-GHz images are consistent with a point source at this resolution, giving an upper limit on the physical size of the source at the galaxy of $\sim 6$ -- $13$\,kpc. Using the  task IMFIT to fit source models to the images, we obtain flux densities of $S_{5.5} = 125.7 \pm 0.4$ and $S_{9.5} = 56.8 \pm 0.9$\,mJy. At the lower frequency end of our observations we also estimate a flux density of $S_{4.85} = 140.6 \pm 0.4$\,mJy, which is consistent with the total flux density of the source $S_{4.85} = 155 \pm 15$\,mJy measured by the Parkes-MIT-NRAO surveys (PMN; \citealt{Wright:1996})
\footnote{\url{https://www.parkes.atnf.csiro.au/observing/databases/pmn/pmn.html}}. The highest resolution 9.5\,GHz image is shown in \autoref{figure:J224500-343030_rgb_image} and is consistent with an offset of 3.4\,arcsec, which at the redshift of the galaxy equates to a physical separation of about 17\,kpc. This confirms that the radio source is not associated with the nucleus and is likely a background AGN. 

\begin{figure*}
\centering
\includegraphics[width=\textwidth]{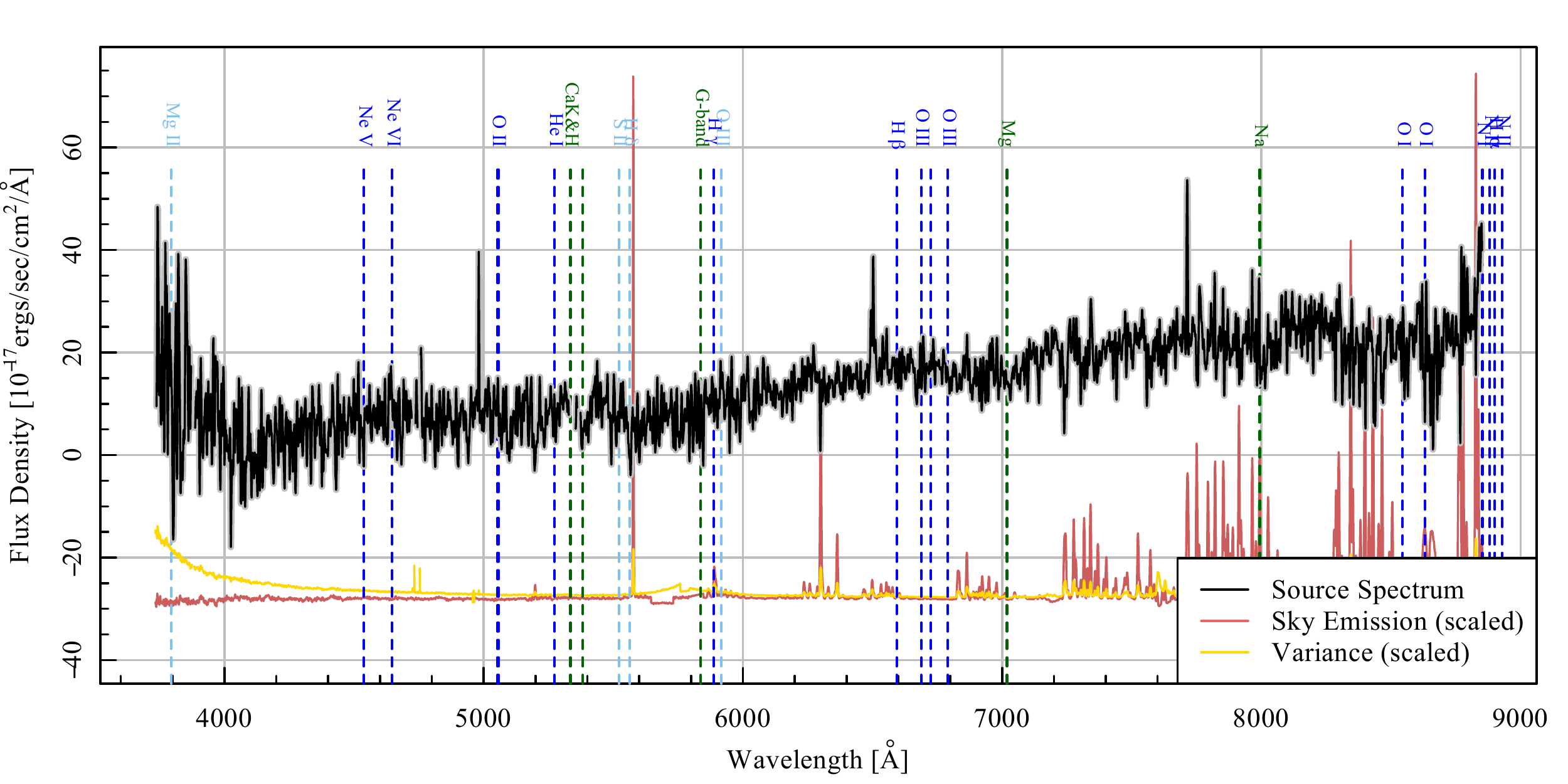}
\caption{Optical spectrum from the central 2.1\,arcsec of the intervening galaxy GAMA\,J224500.05$-$343031.7. The spectrum, smoothed for visual clarity using a 9th-order Hanning window, is consistent with an early-type galaxy at a redshift of $z = 0.3562$. The expected positions of spectral lines are as indicated by the vertical dashed lines; dark blue denotes common emission lines, green denotes common absorption lines and light blue denotes lines commonly associated with AGNs.}
\label{figure:J224500-343030_optical_spectrum}
\end{figure*}

The optical spectrum of GAMA\,J224500.05$-$343031.7, shown in \autoref{figure:J224500-343030_optical_spectrum}, was produced by the GAMA survey using a 2.1-arcsec aperture from the 2dF/AAOmega facility (\citealt{Saunders:2004, Smith:2004b, Sharp:2006}) on the 3.9-m Anglo-Australian Telescope (AAT). Observations were carried out on 2012 November 13 and November 17, with a total exposure time of 4100\,s and a median seeing of 0.7\,arcsec. The original GAMA spectroscopic data for this source were found to have poor spectral reduction due to a strong splicing artefact between the blue and red spectral arm (a known issue in some GAMA spectra, see \citealt{Hopkins:2013}). This artefact does not affect the redshift measurement, which was given a probability match of 99.89\,per\,cent, with a corresponding GAMA quality flag of $nQ = 4$. None the less, for this work the raw spectroscopic data were extracted from the GAMA data base, each arm independently re-flux-calibrated using the $g$- and $r$-band broad-band photometry and the arms recombined; resulting in the spectrum displayed in \autoref{figure:J224500-343030_optical_spectrum}. This re-reduced spectrum was then independently redshifted using the \textsc{AutoZ} code (\citealt{Baldry:2014}), and a high confidence redshift confirmed at $z = 0.3562$. It is characteristic of an early-type galaxy, exhibiting a $4000$\,\AA\, break, Na, Mg, Ca\,{\sc ii} H and K absorption lines, and no detected emission lines. 

\begin{figure*}
\centering
\includegraphics[width=\textwidth]{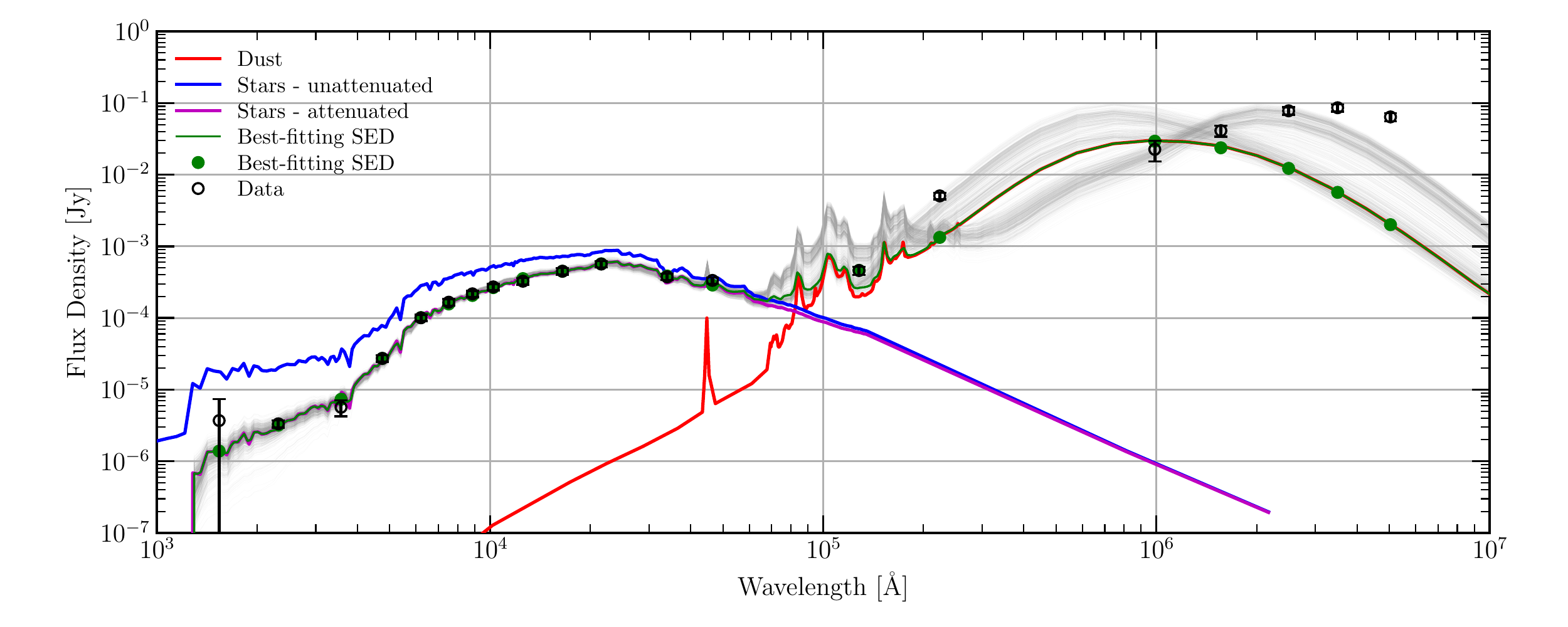}
\caption{The SED of the intervening galaxy GAMA\,J224500.05$-$343031.7. The photometric data shown were compiled using the GAMA Panchromatic Data Release (Bellstedt et al. in preparation). The data (grey points) were extracted from each image using \textsc{ProFound} (\citealt{Robotham:2018}). Template fitting (coloured lines) was carried out using \textsc{ProSpect} (\citealt{Robotham:2020}). The blue line denotes the best-fitting stellar light using the population synthesis models of \citet{Bruzal:2003}. The magenta line is the best-fitting stellar light after attenuating the blue line using the dust model of \citet{Charlot:2000}. The red line is the best-fitting dust emission in the far infrared using the model of \citet{Dale:2014}. Grey lines show 1000 randomly sampled iterations from the MCMC chain, indicating the sampled range of the model SED.}
\label{figure:J224500-343030_SED}
\end{figure*}

In \autoref{figure:J224500-343030_SED} we show the spectral energy distribution of the galaxy using photometric data from the GAMA Panchromatic Data Release (Bellstedt et al. in preparation). SED template fitting was carried out using \textsc{ProSpect} (\citealt{Robotham:2020}), which combines stellar population synthesis models by \cite{Bruzal:2003}, attenuates the resulting stellar light using the dust model of \cite{Charlot:2000} and simultaneously fits the expected dust emission in the far infrared using the model of \cite{Dale:2014}. We parametrize the modelled star formation history as a skewed-normal function, described by four free parameters. These free parameters specify at what age the star formation peaks, the peak star formation rate, the "peakiness" of the star formation history, and the skewness of the star formation history. We do not include a separate star formation burst. We fit for the final metallicity of the galaxy, where the metallicity evolution follows the build-up of stellar mass. 
We estimate a total stellar mass for the galaxy of $M_{\star} = 2.85^{+0.39}_{-0.34} \times 10^{11}$\,M$_{\odot}$ and a star formation rate, averaged over the last 100\,Myr, of SFR $= 6.3^{+21.2}_{-4.3}$  M$_{\odot}$\,yr$^{-1}$. 

The wavelength of the far infrared (FIR) peak of $\lambda \sim 3.5\times10^{6}$\,\AA\, is longer than the typical range of $\lambda \sim (0.7-2.0)\times10^{6}$\,\AA\, for a galaxy at the given redshift, indicating the possibility that the FIR is being contaminated by the background source at a redshift $z \gtrsim 1.4$. To ensure that the potentially-contaminated FIR flux is not impacting the derived stellar mass and SFR, we repeat the SED-fitting without the FIR bands. We obtain values of $M_{\star} = 3.28^{+5.31}_{-0.61}$\,M$_{\odot}$ and SFR $= 3.3^{+55.6}_{-2.5}$  M$_{\odot}$\,yr$^{-1}$, which are consistent within uncertainty with the parameters derived when including the FIR flux, albeit with a significantly larger range. 

The H\,$\alpha$ line is redshifted beyond the observed spectrum, but using the relationship given by \cite{Kewley:2004} we predict from the SFR a luminosity range for the [O\,{\sc ii}]\,$\lambda{3727}$ emission line of $L_{\rm [O\sc{II}]} \sim  (0.12 - 8.9) \times 10^{42}$\,erg\,s$^{-1}$, which at the redshift of the galaxy is equivalent to a flux of $S_{\rm [O\sc{II}]} \sim (2.8 - 210) \times 10^{-16}$\,erg\,s$^{-1}$\,cm$^{-2}$. This should be detectable in the optical spectrum shown in \autoref{figure:J224500-343030_optical_spectrum} and non-detection could either be due to dust obscuration or to a large fraction of the star formation occurring outside of the spectroscopic aperture. From the dust attenuation at $\lambda = 5000$\,\AA\, shown in \autoref{figure:J224500-343030_SED}, it is possible that the line flux density is attenuated by up to 75\,per\,cent. This is potentially sufficient to obscure the expected [O\,{\sc ii}]\,$\lambda{3727}$ emission, particularly at the lower end of the predicted flux range. However, we also note that evidence for concentrated star formation is seen in the colour image in \autoref{figure:J224500-343030_rgb_image} at the edges of the galaxy (blue regions), and the presence of high column densities of cold ($T \sim 100$\,K) \mbox{H\,{\sc i}} gas at a similar galactocentric radius would be consistent with this scenario. 

At a galactocentric radius of 17\,kpc, the integrated optical depth of the absorption line, $\int{\tau_{21}\,\mathrm{d}v} = (0.64 \pm 0.06)$\,km\,s$^{-1}$, is consistent with the spatial distribution of other intervening 21-cm absorbers reported in the literature (see e.g. \citealt{Curran:2016c, Dutta:2017}). If we assume that the spin temperature is greater than or equal to $100$\,K, then the lower limit \mbox{H\,{\sc i}} column density is $N_{\rm HI} \geq (1.2 \pm 0.1) \times 10^{20}$\,cm$^{-2}$, with a corresponding surface density of $\Sigma_{\rm HI} \gtrsim (0.96 \pm 0.1)$\,M$_{\odot}$\,pc$^{-2}$. This is the average value across the unresolved background source, which at the distance of the galaxy has a physical size $d_{\rm src} \lesssim$\,10\,kpc. In their study of \mbox{H\,{\sc i}} 21-cm absorption around galaxies at $z < 0.4$, \cite{Dutta:2017} found that most 21-cm absorbers do not trace the dusty stellar disc, but are aligned with the major axis and so are likely coplanar with any extended \mbox{H\,{\sc i}} disc. If the absorber detected in GAMA\,J224500.05-343031.7 does form part of a larger \mbox{H\,{\sc i}} disc, then  the \mbox{H\,{\sc i}} size-mass relation (see e.g. \citealt{Wang:2016} and references therein) would imply a total \mbox{H\,{\sc i}} mass of $M_{\rm HI} \gtrsim 3 \times 10^{9}$\,M$_{\odot}$. Massive and extended \mbox{H\,{\sc i}} discs are rare in early type galaxies, but have been seen in the nearby Universe (\citealt{Serra:2011}). We may therefore be seeing 21-cm absorption through such an \mbox{H\,{\sc i}} disc in GAMA\,J224500.05$-$343031.7. 

\section{Conclusions}\label{section:summary}
		
Using the ASKAP-12 sub-array of the Australian SKA Pathfinder we have carried out the first widefield spectroscopically blind 21-cm absorption survey at cosmological distances in the GAMA\,23 field, covering redshifts between $z = 0.34$ and 0.79 over a sky area of 50\,$deg^{2}$. 

In a blind search of the data towards 1253 radio sources we did not detect any 21-cm absorbers. For a fiducial spin temperature of $T_{\rm spin} = 100$\,K and a source covering fraction of $c_{\rm f} = 1$, we find that our data are sensitive to intervening 21-cm absorption in DLAs over a comoving absorption path length $\Delta{X} \approx 6.6 \pm 0.3$, and for super-DLAs ($N_{\rm HI} \geq 2 \times 10^{21}$\,cm$^{-2}$) $\Delta{X} \approx 111 \pm 6$. The 95\,per\,cent upper limits on the $N_{\rm HI}$ frequency distribution function, $f(N_{\rm HI}, X)$, are consistent with that measured using 21-cm emission-line surveys at $z \approx 0$ (e.g. \citealt{Zwaan:2005}) and DLA surveys at $z \approx 3$ (e.g. \citealt{Noterdaeme:2009, Noterdaeme:2012}), and with the 21-cm absorption survey at $z \approx 0$ by \cite{Darling:2011}. We calculate the probability of obtaining zero detections as a function of $T_{\rm spin}/c_{\rm f}$ and find that this result is expected with a probability of 64\,per\,cent if the harmonic mean  spin temperature is equal to that in the Milky Way ($T_{\rm spin} \approx 300$\,K; \citealt{Murray:2018}).

We also find that this result is in tension with the detection rates of previous targeted surveys for intrinsic 21-cm absorbers in nearby radio galaxies. If we assume a detection rate equal to 27\,per\,cent, based on the complete sample of \cite{Maccagni:2017}, we find that the probability of detecting no intrinsic absorbers in our data is $\mathrm{Pr}(\mathcal{N}_{\rm abs} = 0) = 0.04$. Given that previous surveys for intrinsic absorbers have selected targets based on their compact flux density, we suggest that the detection rates for future large-scale wide-field surveys are likely to be lower and closer to that found for extended radio galaxies.

By cross-matching the positions of radio sources in our data with optical galaxies that were spectroscopically identified to be within the volume, we were able to search for 21-cm absorption at a lower significance than the Bayes factor threshold used for detection in our blind search. We detected \mbox{H\,{\sc i}} absorption at $z = 0.3563$ towards the radio source NVSS\,J224500$-$343030, later confirming this using more sensitive observations with the full 36-antenna ASKAP. The absorption line is consistent with the spectroscopic redshift ($z = 0.3562$) of GAMA\,J224500.05$-$343031.7, an intervening early-type galaxy at an impact parameter of 17\,kpc. Analysis of the photometric data, available through the GAMA survey of this field, reveals star formation over the past 100\,Myr at the level of SFR $\sim$ 0.5 -- 4\,M$_{\odot}$\,yr$^{-1}$. Visual inspection of the $gri$-band images shows localized star formation in the outskirts of the galaxy at the same galactocentric radius as the absorber. We suggest that the 21-cm absorber, with an \mbox{H\,{\sc i}} column density $N_{\rm HI} \gtrsim 1.2 \times 10^{20}$\,cm$^{-2}$, may form part of a massive \mbox{H\,{\sc i}} disc similar to that seen in some gas-rich early type galaxies in the nearby Universe (\citealt{Serra:2012a}).

Our results demonstrate the feasibility of carrying out a wide-field wide-band survey for 21-cm absorption with ASKAP. The optical spectroscopic coverage of the GAMA23 field allowed us to detect \mbox{H\,{\sc i}} 21-cm absorption below our blind detection threshold. However, we predict that for \mbox{H\,{\sc i}} column densities greater than DLA-like systems ($N_{\rm HI} \geq 2 \times 10^{20}$\,cm$^{-2}$), the planned ASKAP FLASH survey of the southern sky ($\delta < +10\degr$) should be two orders of magnitude more sensitive to the \mbox{H\,{\sc i}} frequency distribution function than these early results, yielding several hundred blind detections of intervening 21-cm absorbers.		
		
\section*{Acknowledgements}

We dedicate this paper to the memory of Richard (Dick) Hunstead, a wonderful colleague, friend and mentor.

We thank Raffaella Morganti and O. Ivy Wong for their helpful comments on an earlier version of the manuscript. We also thank the anonymous referee and the scientific editor for their comments that helped improve this paper.  JRA acknowledges support from a Christ Church Career Development Fellowship. Parts of this research were conducted by the Australian Research Council Centre of Excellence for All-sky Astrophysics in 3D (ASTRO 3D) through project number CE170100013. The National Radio Astronomy Observatory is a facility of the National Science Foundation operated under cooperative agreement by Associated Universities, Inc.

The Australian SKA Pathfinder and the Australia Telescope Compact Array are part of the Australia Telescope National Facility which is managed by CSIRO. Operation of ATCA is funded by the Australian Government. Operation of ASKAP is funded by the Australian Government with support from the National Collaborative Research Infrastructure Strategy. ASKAP uses the resources of the Pawsey Supercomputing Centre. Establishment of ASKAP, the Murchison Radio-astronomy Observatory and the Pawsey Supercomputing Centre are initiatives of the Australian Government, with support from the Government of Western Australia and the Science and Industry Endowment Fund. We acknowledge the Wajarri Yamatji people as the traditional owners of the Observatory site.

The optical images in this work are based on observations made with ESO Telescopes at the La Silla Paranal Observatory under programme IDs 177.A-3016, 177.A-3017, 177.A-3018 and 179.A-2004, and on data products produced by the KiDS consortium. The KiDS production team acknowledges support from: Deutsche Forschungsgemeinschaft, ERC, NOVA and NWO-M grants; Target; the University of Padova, and the University Federico II (Naples).

We have made use of \textsc{Astropy}, a community-developed core PYTHON package for astronomy (\citealt{Astropy:2013}); \textsc{Aplpy}, an open-source plotting package for Python (\citealt{Robitaille:2012}); the NASA/IPAC Extragalactic Database (NED), which is operated by the Jet Propulsion Laboratory, California Institute of Technology, under contract with the National Aeronautics and Space Administration; NASA's Astrophysics Data System Bibliographic Services; and the VizieR catalogue access tool operated at CDS, Strasbourg, France.

%%%%%%%%%%%%%%%%%%%%%%%%%%%%%%%%%%%%%%%%%%%%%%%%%%



%\bibliographystyle{mnras}


\bibliographystyle{mnras}
\bibliography{james}

\begin{thebibliography}{}
\makeatletter
\relax
\def\mn@urlcharsother{\let\do\@makeother \do\$\do\&\do\#\do\^\do\_\do\%\do\~}
\def\mn@doi{\begingroup\mn@urlcharsother \@ifnextchar [ {\mn@doi@}
  {\mn@doi@[]}}
\def\mn@doi@[#1]#2{\def\@tempa{#1}\ifx\@tempa\@empty \href
  {http://dx.doi.org/#2} {doi:#2}\else \href {http://dx.doi.org/#2} {#1}\fi
  \endgroup}
\def\mn@eprint#1#2{\mn@eprint@#1:#2::\@nil}
\def\mn@eprint@arXiv#1{\href {http://arxiv.org/abs/#1} {{\tt arXiv:#1}}}
\def\mn@eprint@dblp#1{\href {http://dblp.uni-trier.de/rec/bibtex/#1.xml}
  {dblp:#1}}
\def\mn@eprint@#1:#2:#3:#4\@nil{\def\@tempa {#1}\def\@tempb {#2}\def\@tempc
  {#3}\ifx \@tempc \@empty \let \@tempc \@tempb \let \@tempb \@tempa \fi \ifx
  \@tempb \@empty \def\@tempb {arXiv}\fi \@ifundefined
  {mn@eprint@\@tempb}{\@tempb:\@tempc}{\expandafter \expandafter \csname
  mn@eprint@\@tempb\endcsname \expandafter{\@tempc}}}

\bibitem[\protect\citeauthoryear{{Aditya} \& {Kanekar}}{{Aditya} \&
  {Kanekar}}{2018}]{Aditya:2018b}
{Aditya} J.~N.~H.~S.,  {Kanekar} N.,  2018, \mn@doi [MNRAS]
  {10.1093/mnras/sty2184}, \href
  {https://ui.adsabs.harvard.edu/abs/2018MNRAS.481.1578A} {481, 1578}

\bibitem[\protect\citeauthoryear{{Aditya}, {Kanekar}  \& {Kurapati}}{{Aditya}
  et~al.}{2016}]{Aditya:2016}
{Aditya} J.~N.~H.~S.,  {Kanekar} N.,   {Kurapati} S.,  2016, \mn@doi [MNRAS]
  {10.1093/mnras/stv2563}, \href
  {https://ui.adsabs.harvard.edu/abs/2016MNRAS.455.4000A} {455, 4000}

\bibitem[\protect\citeauthoryear{{Allison}, {Sadler}  \& {Whiting}}{{Allison}
  et~al.}{2012}]{Allison:2012b}
{Allison} J.~R.,  {Sadler} E.~M.,   {Whiting} M.~T.,  2012, \mn@doi [PASA]
  {10.1071/AS11040}, \href {http://adsabs.harvard.edu/abs/2012PASA...29..221A}
  {29, 221}

\bibitem[\protect\citeauthoryear{{Allison}, {Sadler}  \& {Meekin}}{{Allison}
  et~al.}{2014}]{Allison:2014}
{Allison} J.~R.,  {Sadler} E.~M.,   {Meekin} A.~M.,  2014, \mn@doi [MNRAS]
  {10.1093/mnras/stu289}, \href
  {http://adsabs.harvard.edu/abs/2014MNRAS.440..696A} {440, 696}

\bibitem[\protect\citeauthoryear{{Allison} et~al.,}{{Allison}
  et~al.}{2015}]{Allison:2015}
{Allison} J.~R.,  et~al., 2015, \mn@doi [MNRAS] {10.1093/mnras/stv1532}, \href
  {http://adsabs.harvard.edu/abs/2015MNRAS.453.1249A} {453, 1249}

\bibitem[\protect\citeauthoryear{{Allison} et~al.,}{{Allison}
  et~al.}{2016a}]{Allison:2016a}
{Allison} J.~R.,  et~al., 2016a, \mn@doi [Astronomische Nachrichten]
  {10.1002/asna.201512288}, \href
  {https://ui.adsabs.harvard.edu/abs/2016AN....337..175A} {337, 175}

\bibitem[\protect\citeauthoryear{{Allison}, {Zwaan}, {Duchesne}  \&
  {Curran}}{{Allison} et~al.}{2016b}]{Allison:2016b}
{Allison} J.~R.,  {Zwaan} M.~A.,  {Duchesne} S.~W.,   {Curran} S.~J.,  2016b,
  \mn@doi [MNRAS] {10.1093/mnras/stw1722}, \href
  {http://adsabs.harvard.edu/abs/2016MNRAS.462.1341A} {462, 1341}

\bibitem[\protect\citeauthoryear{{Allison} et~al.,}{{Allison}
  et~al.}{2017}]{Allison:2017}
{Allison} J.~R.,  et~al., 2017, \mn@doi [MNRAS] {10.1093/mnras/stw2860}, \href
  {http://adsabs.harvard.edu/abs/2017MNRAS.465.4450A} {465, 4450}

\bibitem[\protect\citeauthoryear{{Allison} et~al.,}{{Allison}
  et~al.}{2019}]{Allison:2019}
{Allison} J.~R.,  et~al., 2019, \mn@doi [MNRAS] {10.1093/mnras/sty2852}, \href
  {https://ui.adsabs.harvard.edu/abs/2019MNRAS.482.2934A} {482, 2934}

\bibitem[\protect\citeauthoryear{{Astropy Collaboration} et~al.,}{{Astropy
  Collaboration} et~al.}{2013}]{Astropy:2013}
{Astropy Collaboration} et~al., 2013, \mn@doi [A\&A]
  {10.1051/0004-6361/201322068}, \href
  {http://adsabs.harvard.edu/abs/2013A%26A...558A..33A} {558, A33}

\bibitem[\protect\citeauthoryear{{Bahcall} \& {Ekers}}{{Bahcall} \&
  {Ekers}}{1969}]{Bahcall:1969}
{Bahcall} J.~N.,  {Ekers} R.~D.,  1969, \mn@doi [ApJ] {10.1086/150135}, \href
  {http://adsabs.harvard.edu/abs/1969ApJ...157.1055B} {157, 1055}

\bibitem[\protect\citeauthoryear{{Baldry} et~al.,}{{Baldry}
  et~al.}{2014}]{Baldry:2014}
{Baldry} I.~K.,  et~al., 2014, \mn@doi [MNRAS] {10.1093/mnras/stu727}, \href
  {https://ui.adsabs.harvard.edu/abs/2014MNRAS.441.2440B} {441, 2440}

\bibitem[\protect\citeauthoryear{{Becker}, {White}  \& {Helfand}}{{Becker}
  et~al.}{1995}]{Becker:1995}
{Becker} R.~H.,  {White} R.~L.,   {Helfand} D.~J.,  1995, \mn@doi [ApJ]
  {10.1086/176166}, \href {http://adsabs.harvard.edu/abs/1995ApJ...450..559B}
  {450, 559}

\bibitem[\protect\citeauthoryear{{Berg} et~al.,}{{Berg}
  et~al.}{2017}]{Berg:2017}
{Berg} T.~A.~M.,  et~al., 2017, \mn@doi [MNRAS] {10.1093/mnrasl/slw185}, \href
  {https://ui.adsabs.harvard.edu/abs/2017MNRAS.464L..56B} {464, L56}

\bibitem[\protect\citeauthoryear{{Bigiel}, {Leroy}, {Walter}, {Brinks}, {de
  Blok}, {Madore}  \& {Thornley}}{{Bigiel} et~al.}{2008}]{Bigiel:2008}
{Bigiel} F.,  {Leroy} A.,  {Walter} F.,  {Brinks} E.,  {de Blok} W.~J.~G.,
  {Madore} B.,   {Thornley} M.~D.,  2008, \mn@doi [AJ]
  {10.1088/0004-6256/136/6/2846}, \href
  {http://adsabs.harvard.edu/abs/2008AJ....136.2846B} {136, 2846}

\bibitem[\protect\citeauthoryear{{Bird}, {Garnett}  \& {Ho}}{{Bird}
  et~al.}{2017}]{Bird:2017}
{Bird} S.,  {Garnett} R.,   {Ho} S.,  2017, \mn@doi [MNRAS]
  {10.1093/mnras/stw3246}, \href
  {http://adsabs.harvard.edu/abs/2017MNRAS.466.2111B} {466, 2111}

\bibitem[\protect\citeauthoryear{{Bland-Hawthorn}, {Maloney}, {Stephens},
  {Zovaro}  \& {Popping}}{{Bland-Hawthorn} et~al.}{2017}]{Bland-Hawthorn:2017}
{Bland-Hawthorn} J.,  {Maloney} P.~R.,  {Stephens} A.,  {Zovaro} A.,
  {Popping} A.,  2017, \mn@doi [ApJ] {10.3847/1538-4357/aa8f45}, \href
  {http://adsabs.harvard.edu/abs/2017ApJ...849...51B} {849, 51}

\bibitem[\protect\citeauthoryear{{Borthakur}}{{Borthakur}}{2016}]{Borthakur:2016}
{Borthakur} S.,  2016, \mn@doi [ApJ] {10.3847/0004-637X/829/2/128}, \href
  {https://ui.adsabs.harvard.edu/abs/2016ApJ...829..128B} {829, 128}

\bibitem[\protect\citeauthoryear{{Braun}}{{Braun}}{2012}]{Braun:2012}
{Braun} R.,  2012, \mn@doi [ApJ] {10.1088/0004-637X/749/1/87}, \href
  {http://adsabs.harvard.edu/abs/2012ApJ...749...87B} {749, 87}

\bibitem[\protect\citeauthoryear{{Brookes}, {Best}, {Peacock}, {R{\"o}ttgering}
   \& {Dunlop}}{{Brookes} et~al.}{2008}]{Brookes:2008}
{Brookes} M.~H.,  {Best} P.~N.,  {Peacock} J.~A.,  {R{\"o}ttgering} H.~J.~A.,
  {Dunlop} J.~S.,  2008, \mn@doi [MNRAS] {10.1111/j.1365-2966.2008.12786.x},
  \href {http://adsabs.harvard.edu/abs/2008MNRAS.385.1297B} {385, 1297}

\bibitem[\protect\citeauthoryear{{Bruzual} \& {Charlot}}{{Bruzual} \&
  {Charlot}}{2003}]{Bruzal:2003}
{Bruzual} G.,  {Charlot} S.,  2003, \mn@doi [MNRAS]
  {10.1046/j.1365-8711.2003.06897.x}, \href
  {https://ui.adsabs.harvard.edu/abs/2003MNRAS.344.1000B} {344, 1000}

\bibitem[\protect\citeauthoryear{{Carilli} \& {Walter}}{{Carilli} \&
  {Walter}}{2013}]{Carilli:2013}
{Carilli} C.~L.,  {Walter} F.,  2013, \mn@doi [AR\&A]
  {10.1146/annurev-astro-082812-140953}, \href
  {http://adsabs.harvard.edu/abs/2013ARA%26A..51..105C} {51, 105}

\bibitem[\protect\citeauthoryear{{Charlot} \& {Fall}}{{Charlot} \&
  {Fall}}{2000}]{Charlot:2000}
{Charlot} S.,  {Fall} S.~M.,  2000, \mn@doi [ApJ] {10.1086/309250}, \href
  {https://ui.adsabs.harvard.edu/abs/2000ApJ...539..718C} {539, 718}

\bibitem[\protect\citeauthoryear{{Chippendale} et~al.,}{{Chippendale}
  et~al.}{2015}]{Chippendale:2015}
{Chippendale} A.~P.,  et~al., 2015, in 2015 International Conference on
  Electromagnetics in Advanced Applications (ICEAA). IEEE, pp 541 -- 544
  (\mn@eprint {arXiv} {1509.00544})

\bibitem[\protect\citeauthoryear{{Combes}, {Garc{\'{\i}}a-Burillo}, {Braine},
  {Schinnerer}, {Walter}  \& {Colina}}{{Combes} et~al.}{2013}]{Combes:2013}
{Combes} F.,  {Garc{\'{\i}}a-Burillo} S.,  {Braine} J.,  {Schinnerer} E.,
  {Walter} F.,   {Colina} L.,  2013, \mn@doi [A\&A]
  {10.1051/0004-6361/201220392}, \href
  {http://adsabs.harvard.edu/abs/2013A%26A...550A..41C} {550, A41}

\bibitem[\protect\citeauthoryear{{Condon}, {Cotton}, {Greisen}, {Yin},
  {Perley}, {Taylor}  \& {Broderick}}{{Condon} et~al.}{1998}]{Condon:1998}
{Condon} J.~J.,  {Cotton} W.~D.,  {Greisen} E.~W.,  {Yin} Q.~F.,  {Perley}
  R.~A.,  {Taylor} G.~B.,   {Broderick} J.~J.,  1998, \mn@doi [AJ]
  {10.1086/300337}, \href {http://adsabs.harvard.edu/abs/1998AJ....115.1693C}
  {115, 1693}

\bibitem[\protect\citeauthoryear{{Cooke}, {Pettini}  \& {Jorgenson}}{{Cooke}
  et~al.}{2015}]{Cooke:2015}
{Cooke} R.~J.,  {Pettini} M.,   {Jorgenson} R.~A.,  2015, \mn@doi [ApJ]
  {10.1088/0004-637X/800/1/12}, \href
  {https://ui.adsabs.harvard.edu/abs/2015ApJ...800...12C} {800, 12}

\bibitem[\protect\citeauthoryear{{Cornwell} \& {Perley}}{{Cornwell} \&
  {Perley}}{1992}]{Cornwell:1992}
{Cornwell} T.~J.,  {Perley} R.~A.,  1992, A\&A, \href
  {https://ui.adsabs.harvard.edu/abs/1992A&A...261..353C} {261, 353}

\bibitem[\protect\citeauthoryear{{Cornwell}, {Golap}  \&
  {Bhatnagar}}{{Cornwell} et~al.}{2008}]{Cornwell:2008}
{Cornwell} T.~J.,  {Golap} K.,   {Bhatnagar} S.,  2008, \mn@doi [IEEE Journal
  of Selected Topics in Signal Processing] {10.1109/JSTSP.2008.2005290}, \href
  {https://ui.adsabs.harvard.edu/abs/2008ISTSP...2..647C} {2, 647}

\bibitem[\protect\citeauthoryear{{Crighton} et~al.,}{{Crighton}
  et~al.}{2015}]{Crighton:2015}
{Crighton} N.~H.~M.,  et~al., 2015, \mn@doi [MNRAS] {10.1093/mnras/stv1182},
  \href {http://adsabs.harvard.edu/abs/2015MNRAS.452..217C} {452, 217}

\bibitem[\protect\citeauthoryear{{Curran}}{{Curran}}{2017}]{Curran:2017a}
{Curran} S.~J.,  2017, \mn@doi [\mnras] {10.1093/mnras/stx933}, \href
  {https://ui.adsabs.harvard.edu/abs/2017MNRAS.470.3159C} {470, 3159}

\bibitem[\protect\citeauthoryear{{Curran}}{{Curran}}{2019}]{Curran:2019a}
{Curran} S.~J.,  2019, \mn@doi [MNRAS] {10.1093/mnras/stz215}, \href
  {https://ui.adsabs.harvard.edu/abs/2019MNRAS.484.3911C} {484, 3911}

\bibitem[\protect\citeauthoryear{{Curran} \& {Whiting}}{{Curran} \&
  {Whiting}}{2010}]{Curran:2010}
{Curran} S.~J.,  {Whiting} M.~T.,  2010, \mn@doi [ApJ]
  {10.1088/0004-637X/712/1/303}, \href
  {http://adsabs.harvard.edu/abs/2010ApJ...712..303C} {712, 303}

\bibitem[\protect\citeauthoryear{{Curran} \& {Whiting}}{{Curran} \&
  {Whiting}}{2012}]{Curran:2012a}
{Curran} S.~J.,  {Whiting} M.~T.,  2012, \mn@doi [ApJ]
  {10.1088/0004-637X/759/2/117}, \href
  {http://adsabs.harvard.edu/abs/2012ApJ...759..117C} {759, 117}

\bibitem[\protect\citeauthoryear{{Curran}, {Murphy}, {Pihlstr{\"o}m}, {Webb}
  \& {Purcell}}{{Curran} et~al.}{2005}]{Curran:2005}
{Curran} S.~J.,  {Murphy} M.~T.,  {Pihlstr{\"o}m} Y.~M.,  {Webb} J.~K.,
  {Purcell} C.~R.,  2005, \mn@doi [MNRAS] {10.1111/j.1365-2966.2004.08594.x},
  \href {http://adsabs.harvard.edu/abs/2005MNRAS.356.1509C} {356, 1509}

\bibitem[\protect\citeauthoryear{{Curran}, {Whiting}, {Wiklind}, {Webb},
  {Murphy}  \& {Purcell}}{{Curran} et~al.}{2008}]{Curran:2008}
{Curran} S.~J.,  {Whiting} M.~T.,  {Wiklind} T.,  {Webb} J.~K.,  {Murphy}
  M.~T.,   {Purcell} C.~R.,  2008, \mn@doi [MNRAS]
  {10.1111/j.1365-2966.2008.13925.x}, \href
  {http://adsabs.harvard.edu/abs/2008MNRAS.391..765C} {391, 765}

\bibitem[\protect\citeauthoryear{{Curran}, {Allison}, {Glowacki}, {Whiting}  \&
  {Sadler}}{{Curran} et~al.}{2013}]{Curran:2013b}
{Curran} S.~J.,  {Allison} J.~R.,  {Glowacki} M.,  {Whiting} M.~T.,   {Sadler}
  E.~M.,  2013, \mn@doi [MNRAS] {10.1093/mnras/stt438}, \href
  {http://adsabs.harvard.edu/abs/2013MNRAS.431.3408C} {431, 3408}

\bibitem[\protect\citeauthoryear{{Curran}, {Reeves}, {Allison}  \&
  {Sadler}}{{Curran} et~al.}{2016a}]{Curran:2016c}
{Curran} S.~J.,  {Reeves} S.~N.,  {Allison} J.~R.,   {Sadler} E.~M.,  2016a,
  \mn@doi [MNRAS] {10.1093/mnras/stw943}, \href
  {https://ui.adsabs.harvard.edu/abs/2016MNRAS.459.4136C} {459, 4136}

\bibitem[\protect\citeauthoryear{{Curran}, {Duchesne}, {Divoli}  \&
  {Allison}}{{Curran} et~al.}{2016b}]{Curran:2016a}
{Curran} S.~J.,  {Duchesne} S.~W.,  {Divoli} A.,   {Allison} J.~R.,  2016b,
  \mn@doi [MNRAS] {10.1093/mnras/stw1938}, \href
  {http://adsabs.harvard.edu/abs/2016MNRAS.462.4197C} {462, 4197}

\bibitem[\protect\citeauthoryear{{Curran}, {Hunstead}, {Johnston}, {Whiting},
  {Sadler}, {Allison}  \& {Athreya}}{{Curran} et~al.}{2019}]{Curran:2019b}
{Curran} S.~J.,  {Hunstead} R.~W.,  {Johnston} H.~M.,  {Whiting} M.~T.,
  {Sadler} E.~M.,  {Allison} J.~R.,   {Athreya} R.,  2019, \mn@doi [MNRAS]
  {10.1093/mnras/stz038}, \href
  {https://ui.adsabs.harvard.edu/abs/2019MNRAS.484.1182C} {484, 1182}

\bibitem[\protect\citeauthoryear{{Dale}, {Helou}, {Magdis}, {Armus},
  {D{\'\i}az-Santos}  \& {Shi}}{{Dale} et~al.}{2014}]{Dale:2014}
{Dale} D.~A.,  {Helou} G.,  {Magdis} G.~E.,  {Armus} L.,  {D{\'\i}az-Santos}
  T.,   {Shi} Y.,  2014, \mn@doi [ApJ] {10.1088/0004-637X/784/1/83}, \href
  {https://ui.adsabs.harvard.edu/abs/2014ApJ...784...83D} {784, 83}

\bibitem[\protect\citeauthoryear{{Darling}, {Giovanelli}, {Haynes}, {Bolatto}
  \& {Bower}}{{Darling} et~al.}{2004}]{Darling:2004}
{Darling} J.,  {Giovanelli} R.,  {Haynes} M.~P.,  {Bolatto} A.~D.,   {Bower}
  G.~C.,  2004, \mn@doi [ApJ] {10.1086/425143}, \href
  {http://adsabs.harvard.edu/abs/2004ApJ...613L.101D} {613, L101}

\bibitem[\protect\citeauthoryear{{Darling}, {Macdonald}, {Haynes}  \&
  {Giovanelli}}{{Darling} et~al.}{2011}]{Darling:2011}
{Darling} J.,  {Macdonald} E.~P.,  {Haynes} M.~P.,   {Giovanelli} R.,  2011,
  \mn@doi [ApJ] {10.1088/0004-637X/742/1/60}, \href
  {http://adsabs.harvard.edu/abs/2011ApJ...742...60D} {742, 60}

\bibitem[\protect\citeauthoryear{{De Cia}, {Ledoux}, {Petitjean}  \&
  {Savaglio}}{{De Cia} et~al.}{2018}]{DeCia:2018}
{De Cia} A.,  {Ledoux} C.,  {Petitjean} P.,   {Savaglio} S.,  2018, \mn@doi
  [A&A] {10.1051/0004-6361/201731970}, \href
  {https://ui.adsabs.harvard.edu/abs/2018A&A...611A..76D} {611, A76}

\bibitem[\protect\citeauthoryear{{De Zotti}, {Massardi}, {Negrello}  \&
  {Wall}}{{De Zotti} et~al.}{2010}]{deZotti:2010}
{De Zotti} G.,  {Massardi} M.,  {Negrello} M.,   {Wall} J.,  2010, \mn@doi
  [A\&AR] {10.1007/s00159-009-0026-0}, \href
  {http://adsabs.harvard.edu/abs/2010A%26ARv..18....1D} {18, 1}

\bibitem[\protect\citeauthoryear{{Decarli} et~al.,}{{Decarli}
  et~al.}{2016}]{Decarli:2016b}
{Decarli} R.,  et~al., 2016, \mn@doi [ApJ] {10.3847/1538-4357/833/1/69}, \href
  {https://ui.adsabs.harvard.edu/abs/2016ApJ...833...69D} {833, 69}

\bibitem[\protect\citeauthoryear{{Decarli} et~al.,}{{Decarli}
  et~al.}{2019}]{Decarli:2019}
{Decarli} R.,  et~al., 2019, submitted, preprint (arXiv:1903.09164), \href
  {https://ui.adsabs.harvard.edu/abs/2019arXiv190309164D} {p. arXiv:1903.09164}

\bibitem[\protect\citeauthoryear{{Dessauges-Zavadsky}, {Ellison}  \&
  {Murphy}}{{Dessauges-Zavadsky} et~al.}{2009}]{Dessauges-Zavadsky:2009}
{Dessauges-Zavadsky} M.,  {Ellison} S.~L.,   {Murphy} M.~T.,  2009, \mn@doi
  [MNRAS] {10.1111/j.1745-3933.2009.00662.x}, \href
  {https://ui.adsabs.harvard.edu/abs/2009MNRAS.396L..61D} {396, L61}

\bibitem[\protect\citeauthoryear{{Dickey}, {Mebold}, {Marx}, {Amy}, {Haynes}
  \& {Wilson}}{{Dickey} et~al.}{1994}]{Dickey:1994}
{Dickey} J.~M.,  {Mebold} U.,  {Marx} M.,  {Amy} S.,  {Haynes} R.~F.,
  {Wilson} W.,  1994, A\&A, \href
  {http://adsabs.harvard.edu/abs/1994A%26A...289..357D} {289, 357}

\bibitem[\protect\citeauthoryear{{Dickey}, {Mebold}, {Stanimirovic}  \&
  {Staveley-Smith}}{{Dickey} et~al.}{2000}]{Dickey:2000}
{Dickey} J.~M.,  {Mebold} U.,  {Stanimirovic} S.,   {Staveley-Smith} L.,  2000,
  \mn@doi [ApJ] {10.1086/308953}, \href
  {http://adsabs.harvard.edu/abs/2000ApJ...536..756D} {536, 756}

\bibitem[\protect\citeauthoryear{{Driver} et~al.,}{{Driver}
  et~al.}{2018}]{Driver:2018}
{Driver} S.~P.,  et~al., 2018, \mn@doi [MNRAS] {10.1093/mnras/stx2728}, \href
  {https://ui.adsabs.harvard.edu/abs/2018MNRAS.475.2891D} {475, 2891}

\bibitem[\protect\citeauthoryear{{Dutta}, {Srianand}, {Gupta}, {Momjian},
  {Noterdaeme}, {Petitjean}  \& {Rahmani}}{{Dutta} et~al.}{2017}]{Dutta:2017}
{Dutta} R.,  {Srianand} R.,  {Gupta} N.,  {Momjian} E.,  {Noterdaeme} P.,
  {Petitjean} P.,   {Rahmani} H.,  2017, \mn@doi [\mnras]
  {10.1093/mnras/stw2689}, \href
  {https://ui.adsabs.harvard.edu/abs/2017MNRAS.465..588D} {465, 588}

\bibitem[\protect\citeauthoryear{{Ellison} \& {Lopez}}{{Ellison} \&
  {Lopez}}{2009}]{Ellison:2009}
{Ellison} S.~L.,  {Lopez} S.,  2009, \mn@doi [MNRAS]
  {10.1111/j.1365-2966.2009.14947.x}, \href
  {https://ui.adsabs.harvard.edu/abs/2009MNRAS.397..467E} {397, 467}

\bibitem[\protect\citeauthoryear{{Ellison}, {Brown}, {Catinella}  \&
  {Cortese}}{{Ellison} et~al.}{2019}]{Ellison:2019}
{Ellison} S.~L.,  {Brown} T.,  {Catinella} B.,   {Cortese} L.,  2019, \mn@doi
  [MNRAS] {10.1093/mnras/sty3139}, \href
  {https://ui.adsabs.harvard.edu/abs/2019MNRAS.482.5694E} {482, 5694}

\bibitem[\protect\citeauthoryear{{Fern{\'a}ndez} et~al.,}{{Fern{\'a}ndez}
  et~al.}{2016}]{Fernandez:2016}
{Fern{\'a}ndez} X.,  et~al., 2016, \mn@doi [ApJ] {10.3847/2041-8205/824/1/L1},
  \href {http://adsabs.harvard.edu/abs/2016ApJ...824L...1F} {824, L1}

\bibitem[\protect\citeauthoryear{{Feroz} \& {Hobson}}{{Feroz} \&
  {Hobson}}{2008}]{Feroz:2008}
{Feroz} F.,  {Hobson} M.~P.,  2008, \mn@doi [MNRAS]
  {10.1111/j.1365-2966.2007.12353.x}, \href
  {http://adsabs.harvard.edu/abs/2008MNRAS.384..449F} {384, 449}

\bibitem[\protect\citeauthoryear{{Feroz}, {Hobson}  \& {Bridges}}{{Feroz}
  et~al.}{2009}]{Feroz:2009b}
{Feroz} F.,  {Hobson} M.~P.,   {Bridges} M.,  2009, \mn@doi [MNRAS]
  {10.1111/j.1365-2966.2009.14548.x}, \href
  {http://adsabs.harvard.edu/abs/2009MNRAS.398.1601F} {398, 1601}

\bibitem[\protect\citeauthoryear{{Ger{\'e}b}, {Maccagni}, {Morganti}  \&
  {Oosterloo}}{{Ger{\'e}b} et~al.}{2015}]{Gereb:2015}
{Ger{\'e}b} K.,  {Maccagni} F.~M.,  {Morganti} R.,   {Oosterloo} T.~A.,  2015,
  \mn@doi [A\&A] {10.1051/0004-6361/201424655}, \href
  {http://adsabs.harvard.edu/abs/2015A%26A...575A..44G} {575, A44}

\bibitem[\protect\citeauthoryear{{Giovanelli} \& {Haynes}}{{Giovanelli} \&
  {Haynes}}{2016}]{Giovanelli:2016}
{Giovanelli} R.,  {Haynes} M.~P.,  2016, \mn@doi [A\&AR]
  {10.1007/s00159-015-0085-3}, \href
  {http://adsabs.harvard.edu/abs/2016A%26ARv..24....1G} {24, 1}

\bibitem[\protect\citeauthoryear{{Giovanelli} et~al.,}{{Giovanelli}
  et~al.}{2005}]{Giovanelli:2005}
{Giovanelli} R.,  et~al., 2005, \mn@doi [AJ] {10.1086/497431}, \href
  {https://ui.adsabs.harvard.edu/abs/2005AJ....130.2598G} {130, 2598}

\bibitem[\protect\citeauthoryear{{Glowacki} et~al.,}{{Glowacki}
  et~al.}{2019}]{Glowacki:2019}
{Glowacki} M.,  et~al., 2019, \mn@doi [MNRAS] {10.1093/mnras/stz2452}, \href
  {https://ui.adsabs.harvard.edu/abs/2019MNRAS.489.4926G} {489, 4926}

\bibitem[\protect\citeauthoryear{{Grasha}, {Darling}, {Bolatto}, {Leroy}  \&
  {Stocke}}{{Grasha} et~al.}{2019}]{Grasha:2019}
{Grasha} K.,  {Darling} J.,  {Bolatto} A.,  {Leroy} A.~K.,   {Stocke} J.~T.,
  2019, \mn@doi [\apjs] {10.3847/1538-4365/ab4906}, \href
  {https://ui.adsabs.harvard.edu/abs/2019ApJS..245....3G} {245, 3}

\bibitem[\protect\citeauthoryear{{Gupta} et~al.,}{{Gupta}
  et~al.}{2016}]{Gupta:2016}
{Gupta} N.,  et~al., 2016, in Proceedings of MeerKAT Science: On the Pathway to
  the SKA. 25-27 May, 2016 Stellenbosch, South Africa (MeerKAT2016). p.~14
  (\mn@eprint {arXiv} {1708.07371})

\bibitem[\protect\citeauthoryear{{Gupta} et~al.,}{{Gupta}
  et~al.}{2018}]{Gupta:2018}
{Gupta} N.,  et~al., 2018, \mn@doi [MNRAS] {10.1093/mnras/sty384}, \href
  {https://ui.adsabs.harvard.edu/abs/2018MNRAS.476.2432G} {476, 2432}

\bibitem[\protect\citeauthoryear{{Heiles} \& {Troland}}{{Heiles} \&
  {Troland}}{2003}]{Heiles:2003b}
{Heiles} C.,  {Troland} T.~H.,  2003, \mn@doi [ApJ] {10.1086/367828}, \href
  {http://adsabs.harvard.edu/abs/2003ApJ...586.1067H} {586, 1067}

\bibitem[\protect\citeauthoryear{{Helfand}, {White}  \& {Becker}}{{Helfand}
  et~al.}{2015}]{Helfand:2015}
{Helfand} D.~J.,  {White} R.~L.,   {Becker} R.~H.,  2015, \mn@doi [ApJ]
  {10.1088/0004-637X/801/1/26}, \href
  {https://ui.adsabs.harvard.edu/abs/2015ApJ...801...26H} {801, 26}

\bibitem[\protect\citeauthoryear{Hellwig, Vessot, Levine, W.~Zitzewitz,
  W.~Allan  \& J.~Glaze}{Hellwig et~al.}{1970}]{Hellwig:1970}
Hellwig H.,  Vessot R.,  Levine M.,  W.~Zitzewitz P.,  W.~Allan D.,   J.~Glaze
  D.,  1970, \mn@doi [Instrumentation and Measurement, IEEE Transactions on]
  {10.1109/TIM.1970.4313902}, 19, 200

\bibitem[\protect\citeauthoryear{{Hopkins} \& {Beacom}}{{Hopkins} \&
  {Beacom}}{2006}]{Hopkins:2006b}
{Hopkins} A.~M.,  {Beacom} J.~F.,  2006, \mn@doi [ApJ] {10.1086/506610}, \href
  {http://adsabs.harvard.edu/abs/2006ApJ...651..142H} {651, 142}

\bibitem[\protect\citeauthoryear{{Hopkins} et~al.,}{{Hopkins}
  et~al.}{2013}]{Hopkins:2013}
{Hopkins} A.~M.,  et~al., 2013, \mn@doi [MNRAS] {10.1093/mnras/stt030}, \href
  {https://ui.adsabs.harvard.edu/abs/2013MNRAS.430.2047H} {430, 2047}

\bibitem[\protect\citeauthoryear{{Hurley-Walker} et~al.,}{{Hurley-Walker}
  et~al.}{2017}]{Hurley-Walker:2017}
{Hurley-Walker} N.,  et~al., 2017, \mn@doi [MNRAS] {10.1093/mnras/stw2337},
  \href {http://adsabs.harvard.edu/abs/2017MNRAS.464.1146H} {464, 1146}

\bibitem[\protect\citeauthoryear{{Intema}, {Jagannathan}, {Mooley}  \&
  {Frail}}{{Intema} et~al.}{2017}]{Intema:2017}
{Intema} H.~T.,  {Jagannathan} P.,  {Mooley} K.~P.,   {Frail} D.~A.,  2017,
  \mn@doi [A\&A] {10.1051/0004-6361/201628536}, \href
  {http://adsabs.harvard.edu/abs/2017A%26A...598A..78I} {598, A78}

\bibitem[\protect\citeauthoryear{{Isbell}, {Xue}  \& {Fu}}{{Isbell}
  et~al.}{2018}]{Isbell:2018}
{Isbell} J.~W.,  {Xue} R.,   {Fu} H.,  2018, \mn@doi [ApJL]
  {10.3847/2041-8213/aaf872}, \href
  {http://adsabs.harvard.edu/abs/2018ApJ...869L..37I} {869, L37}

\bibitem[\protect\citeauthoryear{{Johnston} et~al.,}{{Johnston}
  et~al.}{2007}]{Johnston:2007}
{Johnston} S.,  et~al., 2007, \mn@doi [PASA] {10.1071/AS07033}, \href
  {http://adsabs.harvard.edu/abs/2007PASA...24..174J} {24, 174}

\bibitem[\protect\citeauthoryear{{Jones}, {Haynes}, {Giovanelli}  \&
  {Moorman}}{{Jones} et~al.}{2018}]{Jones:2018}
{Jones} M.~G.,  {Haynes} M.~P.,  {Giovanelli} R.,   {Moorman} C.,  2018,
  \mn@doi [MNRAS] {10.1093/mnras/sty521}, \href
  {https://ui.adsabs.harvard.edu/abs/2018MNRAS.477....2J} {477, 2}

\bibitem[\protect\citeauthoryear{{Kanekar} \& {Briggs}}{{Kanekar} \&
  {Briggs}}{2004}]{Kanekar:2004}
{Kanekar} N.,  {Briggs} F.~H.,  2004, \mn@doi [New Astron. Rev.]
  {10.1016/j.newar.2004.09.030}, \href
  {http://adsabs.harvard.edu/abs/2004NewAR..48.1259K} {48, 1259}

\bibitem[\protect\citeauthoryear{{Kanekar} et~al.,}{{Kanekar}
  et~al.}{2014}]{Kanekar:2014a}
{Kanekar} N.,  et~al., 2014, \mn@doi [MNRAS] {10.1093/mnras/stt2338}, \href
  {http://adsabs.harvard.edu/abs/2014MNRAS.438.2131K} {438, 2131}

\bibitem[\protect\citeauthoryear{Kass \& Raftery}{Kass \&
  Raftery}{1995}]{Kass:1995}
Kass R.~E.,  Raftery A.~E.,  1995, \mn@doi [Journal of the American Statistical
  Association] {10.1080/01621459.1995.10476572}, 90, 773

\bibitem[\protect\citeauthoryear{{Kauffmann} \& {Heckman}}{{Kauffmann} \&
  {Heckman}}{2009}]{Kauffmann:2009}
{Kauffmann} G.,  {Heckman} T.~M.,  2009, \mn@doi [MNRAS]
  {10.1111/j.1365-2966.2009.14960.x}, \href
  {http://adsabs.harvard.edu/abs/2009MNRAS.397..135K} {397, 135}

\bibitem[\protect\citeauthoryear{{Kauffmann} et~al.,}{{Kauffmann}
  et~al.}{2003}]{Kauffmann:2003}
{Kauffmann} G.,  et~al., 2003, \mn@doi [MNRAS]
  {10.1111/j.1365-2966.2003.07154.x}, \href
  {http://adsabs.harvard.edu/abs/2003MNRAS.346.1055K} {346, 1055}

\bibitem[\protect\citeauthoryear{{Kauffmann} et~al.,}{{Kauffmann}
  et~al.}{2007}]{Kauffmann:2007}
{Kauffmann} G.,  et~al., 2007, \mn@doi [ApJS] {10.1086/516647}, \href
  {http://adsabs.harvard.edu/abs/2007ApJS..173..357K} {173, 357}

\bibitem[\protect\citeauthoryear{{Kennicutt}}{{Kennicutt}}{1998}]{Kennicutt:1998}
{Kennicutt} Jr. R.~C.,  1998, \mn@doi [ApJ] {10.1086/305588}, \href
  {http://adsabs.harvard.edu/abs/1998ApJ...498..541K} {498, 541}

\bibitem[\protect\citeauthoryear{{Kewley}, {Geller}  \& {Jansen}}{{Kewley}
  et~al.}{2004}]{Kewley:2004}
{Kewley} L.~J.,  {Geller} M.~J.,   {Jansen} R.~A.,  2004, \mn@doi [AJ]
  {10.1086/382723}, \href {http://adsabs.harvard.edu/abs/2004AJ....127.2002K}
  {127, 2002}

\bibitem[\protect\citeauthoryear{{Kleiner} et~al.,}{{Kleiner}
  et~al.}{2019}]{Kleiner:2019}
{Kleiner} D.,  et~al., 2019, \mn@doi [MNRAS] {10.1093/mnras/stz2063}, \href
  {https://ui.adsabs.harvard.edu/abs/2019MNRAS.488.5352K} {488, 5352}

\bibitem[\protect\citeauthoryear{{Kuijken} et~al.,}{{Kuijken}
  et~al.}{2019}]{Kuijken:2019}
{Kuijken} K.,  et~al., 2019, \mn@doi [A\&A] {10.1051/0004-6361/201834918},
  \href {https://ui.adsabs.harvard.edu/abs/2019A&A...625A...2K} {625, A2}

\bibitem[\protect\citeauthoryear{{LaMassa}, {Heckman}, {Ptak}  \&
  {Urry}}{{LaMassa} et~al.}{2013}]{LaMassa:2013}
{LaMassa} S.~M.,  {Heckman} T.~M.,  {Ptak} A.,   {Urry} C.~M.,  2013, \mn@doi
  [ApJL] {10.1088/2041-8205/765/2/L33}, \href
  {http://adsabs.harvard.edu/abs/2013ApJ...765L..33L} {765, L33}

\bibitem[\protect\citeauthoryear{{Large}, {Mills}, {Little}, {Crawford}  \&
  {Sutton}}{{Large} et~al.}{1981}]{Large:1981}
{Large} M.~I.,  {Mills} B.~Y.,  {Little} A.~G.,  {Crawford} D.~F.,   {Sutton}
  J.~M.,  1981, MNRAS, \href
  {http://adsabs.harvard.edu/abs/1981MNRAS.194..693L} {194, 693}

\bibitem[\protect\citeauthoryear{{Leahy} et~al.,}{{Leahy}
  et~al.}{2019}]{Leahy:2019}
{Leahy} D.~A.,  et~al., 2019, \mn@doi [PASA] {10.1017/pasa.2019.16}, \href
  {https://ui.adsabs.harvard.edu/abs/2019PASA...36...24L} {36, e024}

\bibitem[\protect\citeauthoryear{{Liske} et~al.,}{{Liske}
  et~al.}{2015}]{Liske:2015}
{Liske} J.,  et~al., 2015, \mn@doi [MNRAS] {10.1093/mnras/stv1436}, \href
  {http://adsabs.harvard.edu/abs/2015MNRAS.452.2087L} {452, 2087}

\bibitem[\protect\citeauthoryear{{Maccagni}, {Morganti}, {Oosterloo},
  {Ger{\'e}b}  \& {Maddox}}{{Maccagni} et~al.}{2017}]{Maccagni:2017}
{Maccagni} F.~M.,  {Morganti} R.,  {Oosterloo} T.~A.,  {Ger{\'e}b} K.,
  {Maddox} N.,  2017, \mn@doi [A\&A] {10.1051/0004-6361/201730563}, \href
  {http://adsabs.harvard.edu/abs/2017A%26A...604A..43M} {604, A43}

\bibitem[\protect\citeauthoryear{{Madau} \& {Dickinson}}{{Madau} \&
  {Dickinson}}{2014}]{Madau:2014}
{Madau} P.,  {Dickinson} M.,  2014, \mn@doi [ARA&A]
  {10.1146/annurev-astro-081811-125615}, \href
  {https://ui.adsabs.harvard.edu/abs/2014ARA&A..52..415M} {52, 415}

\bibitem[\protect\citeauthoryear{{Magdis} et~al.,}{{Magdis}
  et~al.}{2014}]{Magdis:2014}
{Magdis} G.~E.,  et~al., 2014, \mn@doi [ApJ] {10.1088/0004-637X/796/1/63},
  \href {https://ui.adsabs.harvard.edu/abs/2014ApJ...796...63M} {796, 63}

\bibitem[\protect\citeauthoryear{{Martin}, {Papastergis}, {Giovanelli},
  {Haynes}, {Springob}  \& {Stierwalt}}{{Martin} et~al.}{2010}]{Martin:2010}
{Martin} A.~M.,  {Papastergis} E.,  {Giovanelli} R.,  {Haynes} M.~P.,
  {Springob} C.~M.,   {Stierwalt} S.,  2010, \mn@doi [ApJ]
  {10.1088/0004-637X/723/2/1359}, \href
  {http://adsabs.harvard.edu/abs/2010ApJ...723.1359M} {723, 1359}

\bibitem[\protect\citeauthoryear{{Mauch}, {Murphy}, {Buttery}, {Curran},
  {Hunstead}, {Piestrzynski}, {Robertson}  \& {Sadler}}{{Mauch}
  et~al.}{2003}]{Mauch:2003}
{Mauch} T.,  {Murphy} T.,  {Buttery} H.~J.,  {Curran} J.,  {Hunstead} R.~W.,
  {Piestrzynski} B.,  {Robertson} J.~G.,   {Sadler} E.~M.,  2003, \mn@doi
  [MNRAS] {10.1046/j.1365-8711.2003.06605.x}, \href
  {http://adsabs.harvard.edu/abs/2003MNRAS.342.1117M} {342, 1117}

\bibitem[\protect\citeauthoryear{{McMullin}, {Waters}, {Schiebel}, {Young}  \&
  {Golap}}{{McMullin} et~al.}{2007}]{McMullin:2007}
{McMullin} J.~P.,  {Waters} B.,  {Schiebel} D.,  {Young} W.,   {Golap} K.,
  2007, in {Shaw} R.~A.,  {Hill} F.,   {Bell} D.~J.,  eds,  ASP Conf. Ser. Vol.
  376, Astronomical Data Analysis Software and Systems XVI. Astron. Soc. Pac.,
  San Francisco, p.~127

\bibitem[\protect\citeauthoryear{{Meyer}, {Robotham}, {Obreschkow},
  {Westmeier}, {Duffy}  \& {Staveley-Smith}}{{Meyer} et~al.}{2017}]{Meyer:2017}
{Meyer} M.,  {Robotham} A.,  {Obreschkow} D.,  {Westmeier} T.,  {Duffy} A.~R.,
   {Staveley-Smith} L.,  2017, \mn@doi [Publ. Astron. Soc. Aust.]
  {10.1017/pasa.2017.31}, \href
  {https://ui.adsabs.harvard.edu/abs/2017PASA...34...52M} {34}

\bibitem[\protect\citeauthoryear{{Morganti} \& {Oosterloo}}{{Morganti} \&
  {Oosterloo}}{2018}]{Morganti:2018}
{Morganti} R.,  {Oosterloo} T.,  2018, preprint, \href
  {http://adsabs.harvard.edu/abs/2018arXiv180701475M} {} (\mn@eprint {arXiv}
  {1807.01475})

\bibitem[\protect\citeauthoryear{{Morganti}, {Sadler}  \& {Curran}}{{Morganti}
  et~al.}{2015}]{Morganti:2015}
{Morganti} R.,  {Sadler} E.~M.,   {Curran} S.,  2015, in {Bourke} T.,  {Braun}
  R.,  {Fender} R.~P.,   {et al.} eds, Advancing Astrophysics with the Square
  Kilometre Array (AASKA14). Proc. Sci., p.~134 (\mn@eprint {arXiv}
  {1501.01091})

\bibitem[\protect\citeauthoryear{{Moss} et~al.,}{{Moss}
  et~al.}{2017}]{Moss:2017}
{Moss} V.~A.,  et~al., 2017, \mn@doi [MNRAS] {10.1093/mnras/stx1679}, \href
  {http://adsabs.harvard.edu/abs/2017MNRAS.471.2952M} {471, 2952}

\bibitem[\protect\citeauthoryear{{Murray}, {Stanimirovi{\'c}}, {Goss},
  {Heiles}, {Dickey}, {Babler}  \& {Kim}}{{Murray} et~al.}{2018}]{Murray:2018}
{Murray} C.~E.,  {Stanimirovi{\'c}} S.,  {Goss} W.~M.,  {Heiles} C.,  {Dickey}
  J.~M.,  {Babler} B.,   {Kim} C.-G.,  2018, \mn@doi [ApJS]
  {10.3847/1538-4365/aad81a}, \href
  {http://adsabs.harvard.edu/abs/2018ApJS..238...14M} {238, 14}

\bibitem[\protect\citeauthoryear{{Neeleman}, {Prochaska}, {Ribaudo}, {Lehner},
  {Howk}, {Rafelski}  \& {Kanekar}}{{Neeleman} et~al.}{2016}]{Neeleman:2016}
{Neeleman} M.,  {Prochaska} J.~X.,  {Ribaudo} J.,  {Lehner} N.,  {Howk} J.~C.,
  {Rafelski} M.,   {Kanekar} N.,  2016, \mn@doi [ApJ]
  {10.3847/0004-637X/818/2/113}, \href
  {http://adsabs.harvard.edu/abs/2016ApJ...818..113N} {818, 113}

\bibitem[\protect\citeauthoryear{{Norris} et~al.,}{{Norris}
  et~al.}{2011}]{Norris:2011}
{Norris} R.~P.,  et~al., 2011, \mn@doi [Publ. Astron. Soc. Aust.]
  {10.1071/AS11021}, \href {http://adsabs.harvard.edu/abs/2011PASA...28..215N}
  {28, 215}

\bibitem[\protect\citeauthoryear{{Noterdaeme}, {Petitjean}, {Ledoux}  \&
  {Srianand}}{{Noterdaeme} et~al.}{2009}]{Noterdaeme:2009}
{Noterdaeme} P.,  {Petitjean} P.,  {Ledoux} C.,   {Srianand} R.,  2009, \mn@doi
  [A\&A] {10.1051/0004-6361/200912768}, \href
  {http://adsabs.harvard.edu/abs/2009A%26A...505.1087N} {505, 1087}

\bibitem[\protect\citeauthoryear{{Noterdaeme} et~al.,}{{Noterdaeme}
  et~al.}{2012}]{Noterdaeme:2012}
{Noterdaeme} P.,  et~al., 2012, \mn@doi [A\&A] {10.1051/0004-6361/201220259},
  \href {http://adsabs.harvard.edu/abs/2012A%26A...547L...1N} {547, L1}

\bibitem[\protect\citeauthoryear{{Oosterloo}, {Verheijen}, {van Cappellen},
  {Bakker}, {Heald}  \& {Ivashina}}{{Oosterloo} et~al.}{2009}]{Oosterloo:2009}
{Oosterloo} T.,  {Verheijen} M.~A.~W.,  {van Cappellen} W.,  {Bakker} L.,
  {Heald} G.,   {Ivashina} M.,  2009, in {Torchinsky} S.~A.,  {van Ardenne} A.,
   {van den Brink-Havinga} T.,  {van Es} A.~J.~J.,   {Faulkner} A.~J.,  eds,
  Wide Field Astronomy and Technology for the Square Kilometre Array. Proc.
  Sci., p.~70 (\mn@eprint {arXiv} {0912.0093})

\bibitem[\protect\citeauthoryear{{P{\'e}roux}, {Deharveng}, {Le Brun}  \&
  {Cristiani}}{{P{\'e}roux} et~al.}{2004}]{Peroux:2004}
{P{\'e}roux} C.,  {Deharveng} J.-M.,  {Le Brun} V.,   {Cristiani} S.,  2004,
  \mn@doi [MNRAS] {10.1111/j.1365-2966.2004.08018.x}, \href
  {https://ui.adsabs.harvard.edu/abs/2004MNRAS.352.1291P} {352, 1291}

\bibitem[\protect\citeauthoryear{{Pihlstr{\"o}m}, {Conway}  \&
  {Vermeulen}}{{Pihlstr{\"o}m} et~al.}{2003}]{Pihlstrom:2003}
{Pihlstr{\"o}m} Y.~M.,  {Conway} J.~E.,   {Vermeulen} R.~C.,  2003, \mn@doi
  [A\&A] {10.1051/0004-6361:20030469}, \href
  {http://adsabs.harvard.edu/abs/2003A26A...404..871P} {404, 871}

\bibitem[\protect\citeauthoryear{{Rafelski}, {Wolfe}, {Prochaska}, {Neeleman}
  \& {Mendez}}{{Rafelski} et~al.}{2012}]{Rafelski:2012}
{Rafelski} M.,  {Wolfe} A.~M.,  {Prochaska} J.~X.,  {Neeleman} M.,   {Mendez}
  A.~J.,  2012, \mn@doi [ApJ] {10.1088/0004-637X/755/2/89}, \href
  {https://ui.adsabs.harvard.edu/abs/2012ApJ...755...89R} {755, 89}

\bibitem[\protect\citeauthoryear{{Rao}, {Turnshek}  \& {Nestor}}{{Rao}
  et~al.}{2006}]{Rao:2006}
{Rao} S.~M.,  {Turnshek} D.~A.,   {Nestor} D.~B.,  2006, \mn@doi [ApJ]
  {10.1086/498132}, \href {http://adsabs.harvard.edu/abs/2006ApJ...636..610R}
  {636, 610}

\bibitem[\protect\citeauthoryear{{Rao}, {Turnshek}, {Sardane}  \&
  {Monier}}{{Rao} et~al.}{2017}]{Rao:2017}
{Rao} S.~M.,  {Turnshek} D.~A.,  {Sardane} G.~M.,   {Monier} E.~M.,  2017,
  \mn@doi [MNRAS] {10.1093/mnras/stx1787}, \href
  {https://ui.adsabs.harvard.edu/abs/2017MNRAS.471.3428R} {471, 3428}

\bibitem[\protect\citeauthoryear{{Reeves} et~al.,}{{Reeves}
  et~al.}{2016}]{Reeves:2016}
{Reeves} S.~N.,  et~al., 2016, \mn@doi [MNRAS] {10.1093/mnras/stv3011}, \href
  {https://ui.adsabs.harvard.edu/abs/2016MNRAS.457.2613R} {457, 2613}

\bibitem[\protect\citeauthoryear{{Reynolds}}{{Reynolds}}{1994}]{Reynolds:1994}
{Reynolds} J.,  1994, AT Technical Document AT/39.3/040

\bibitem[\protect\citeauthoryear{{Rhee}, {Lah}, {Briggs}, {Chengalur},
  {Colless}, {Willner}, {Ashby}  \& {Le F{\`e}vre}}{{Rhee}
  et~al.}{2018}]{Rhee:2018}
{Rhee} J.,  {Lah} P.,  {Briggs} F.~H.,  {Chengalur} J.~N.,  {Colless} M.,
  {Willner} S.~P.,  {Ashby} M.~L.~N.,   {Le F{\`e}vre} O.,  2018, \mn@doi
  [MNRAS] {10.1093/mnras/stx2461}, \href
  {http://adsabs.harvard.edu/abs/2018MNRAS.473.1879R} {473, 1879}

\bibitem[\protect\citeauthoryear{{Robitaille} \& {Bressert}}{{Robitaille} \&
  {Bressert}}{2012}]{Robitaille:2012}
{Robitaille} T.,  {Bressert} E.,  2012, {APLpy: Astronomical Plotting Library
  in Python}, Astrophysics Source Code Library (\mn@eprint {ascl} {1208.017})

\bibitem[\protect\citeauthoryear{{Robotham}, {Davies}, {Driver}, {Koushan},
  {Taranu}, {Casura}  \& {Liske}}{{Robotham} et~al.}{2018}]{Robotham:2018}
{Robotham} A.~S.~G.,  {Davies} L.~J.~M.,  {Driver} S.~P.,  {Koushan} S.,
  {Taranu} D.~S.,  {Casura} S.,   {Liske} J.,  2018, \mn@doi [MNRAS]
  {10.1093/mnras/sty440}, \href
  {https://ui.adsabs.harvard.edu/abs/2018MNRAS.476.3137R} {476, 3137}

\bibitem[\protect\citeauthoryear{{Robotham}, {Bellstedt}, {Lagos}, {Thorne},
  {Davies}, {Driver}  \& {Bravo}}{{Robotham} et~al.}{2020}]{Robotham:2020}
{Robotham} A.~S.~G.,  {Bellstedt} S.,  {Lagos} C. d.~P.,  {Thorne} J.~E.,
  {Davies} L.~J.,  {Driver} S.~P.,   {Bravo} M.,  2020, submitted, preprint
  (arXiv:2002.06980), \href
  {https://ui.adsabs.harvard.edu/abs/2020arXiv200206980R} {p. arXiv:2002.06980}

\bibitem[\protect\citeauthoryear{{Roy}, {Kanekar}  \& {Chengalur}}{{Roy}
  et~al.}{2013}]{Roy:2013b}
{Roy} N.,  {Kanekar} N.,   {Chengalur} J.~N.,  2013, \mn@doi [MNRAS]
  {10.1093/mnras/stt1746}, \href
  {http://adsabs.harvard.edu/abs/2013MNRAS.436.2366R} {436, 2366}

\bibitem[\protect\citeauthoryear{{S{\'a}nchez-Ram{\'{\i}}rez}
  et~al.,}{{S{\'a}nchez-Ram{\'{\i}}rez} et~al.}{2016}]{Sanchez-Ramirez:2016}
{S{\'a}nchez-Ram{\'{\i}}rez} R.,  et~al., 2016, \mn@doi [MNRAS]
  {10.1093/mnras/stv2732}, \href
  {http://adsabs.harvard.edu/abs/2016MNRAS.456.4488S} {456, 4488}

\bibitem[\protect\citeauthoryear{{Sault}, {Teuben}  \& {Wright}}{{Sault}
  et~al.}{1995}]{Sault:1995}
{Sault} R.~J.,  {Teuben} P.~J.,   {Wright} M.~C.~H.,  1995, in {Shaw} R.~A.,
  {Payne} H.~E.,   {Hayes} J.~J.~E.,  eds,  ASP Conf. Ser. Vol. 77,
  Astronomical Data Analysis Software and Systems IV. Astron. Soc. Pac., San
  Francisco, p.~433 (\mn@eprint {} {arXiv:astro-ph/0612759})

\bibitem[\protect\citeauthoryear{{Saunders} et~al.,}{{Saunders}
  et~al.}{2004}]{Saunders:2004}
{Saunders} W.,  et~al., 2004, in {Moorwood} A. F.~M.,  {Iye} M.,  eds,  Society
  of Photo-Optical Instrumentation Engineers (SPIE) Conference Series Vol.
  5492, \procspie. pp 389--400, \mn@doi{10.1117/12.550871}

\bibitem[\protect\citeauthoryear{{Schmidt}}{{Schmidt}}{1959}]{Schmidt:1959}
{Schmidt} M.,  1959, \mn@doi [ApJ] {10.1086/146614}, \href
  {http://adsabs.harvard.edu/abs/1959ApJ...129..243S} {129, 243}

\bibitem[\protect\citeauthoryear{{Serra}, {Jurek}  \& {Fl{\"o}er}}{{Serra}
  et~al.}{2012a}]{Serra:2012b}
{Serra} P.,  {Jurek} R.,   {Fl{\"o}er} L.,  2012a, \mn@doi [Publ. Astron. Soc.
  Aust.] {10.1071/AS11065}, \href
  {http://adsabs.harvard.edu/abs/2012PASA...29..296S} {29, 296}

\bibitem[\protect\citeauthoryear{{Serra} et~al.,}{{Serra}
  et~al.}{2012b}]{Serra:2012a}
{Serra} P.,  et~al., 2012b, \mn@doi [MNRAS] {10.1111/j.1365-2966.2012.20219.x},
  \href {http://adsabs.harvard.edu/abs/2012MNRAS.422.1835S} {422, 1835}

\bibitem[\protect\citeauthoryear{{Serra} et~al.,}{{Serra}
  et~al.}{2012c}]{Serra:2011}
{Serra} P.,  et~al., 2012c, \mn@doi [\mnras]
  {10.1111/j.1365-2966.2012.20219.x}, \href
  {https://ui.adsabs.harvard.edu/abs/2012MNRAS.422.1835S} {422, 1835}

\bibitem[\protect\citeauthoryear{{Shankar}, {Weinberg}  \&
  {Miralda-Escud{\'e}}}{{Shankar} et~al.}{2009}]{Shankar:2009}
{Shankar} F.,  {Weinberg} D.~H.,   {Miralda-Escud{\'e}} J.,  2009, \mn@doi
  [ApJ] {10.1088/0004-637X/690/1/20}, \href
  {http://adsabs.harvard.edu/abs/2009ApJ...690...20S} {690, 20}

\bibitem[\protect\citeauthoryear{{Sharp} et~al.,}{{Sharp}
  et~al.}{2006}]{Sharp:2006}
{Sharp} R.,  et~al., 2006, in \procspie. p. 62690G (\mn@eprint {arXiv}
  {astro-ph/0606137}), \mn@doi{10.1117/12.671022}

\bibitem[\protect\citeauthoryear{{Shimmins} \& {Bolton}}{{Shimmins} \&
  {Bolton}}{1974}]{Shimmins:1974}
{Shimmins} A.~J.,  {Bolton} J.~G.,  1974, Australian Journal of Physics
  Astrophysical Supplement, \href
  {http://adsabs.harvard.edu/abs/1974AuJPA..32....1S} {32, 1}

\bibitem[\protect\citeauthoryear{{Smith} et~al.,}{{Smith}
  et~al.}{2004}]{Smith:2004b}
{Smith} G.~A.,  et~al., 2004, in {Moorwood} A. F.~M.,  {Iye} M.,  eds,  Society
  of Photo-Optical Instrumentation Engineers (SPIE) Conference Series Vol.
  5492, \procspie. pp 410--420, \mn@doi{10.1117/12.551013}

\bibitem[\protect\citeauthoryear{{Snellen}, {Schilizzi}, {Miley}, {de Bruyn},
  {Bremer}  \& {R{\"o}ttgering}}{{Snellen} et~al.}{2000}]{Snellen:2000a}
{Snellen} I.~A.~G.,  {Schilizzi} R.~T.,  {Miley} G.~K.,  {de Bruyn} A.~G.,
  {Bremer} M.~N.,   {R{\"o}ttgering} H.~J.~A.,  2000, \mn@doi [MNRAS]
  {10.1046/j.1365-8711.2000.03935.x}, \href
  {http://adsabs.harvard.edu/abs/2000MNRAS.319..445S} {319, 445}

\bibitem[\protect\citeauthoryear{{Spergel} et~al.,}{{Spergel}
  et~al.}{2007}]{Spergel:2007}
{Spergel} D.~N.,  et~al., 2007, \mn@doi [ApJS] {10.1086/513700}, \href
  {http://adsabs.harvard.edu/abs/2007ApJS..170..377S} {170, 377}

\bibitem[\protect\citeauthoryear{{Tacconi} et~al.,}{{Tacconi}
  et~al.}{2013}]{Tacconi:2013}
{Tacconi} L.~J.,  et~al., 2013, \mn@doi [ApJ] {10.1088/0004-637X/768/1/74},
  \href {https://ui.adsabs.harvard.edu/abs/2013ApJ...768...74T} {768, 74}

\bibitem[\protect\citeauthoryear{{Ueda}, {Akiyama}, {Ohta}  \& {Miyaji}}{{Ueda}
  et~al.}{2003}]{Ueda:2003}
{Ueda} Y.,  {Akiyama} M.,  {Ohta} K.,   {Miyaji} T.,  2003, \mn@doi [ApJ]
  {10.1086/378940}, \href {http://adsabs.harvard.edu/abs/2003ApJ...598..886U}
  {598, 886}

\bibitem[\protect\citeauthoryear{{Villanueva} et~al.,}{{Villanueva}
  et~al.}{2017}]{Villaneuva:2017}
{Villanueva} V.,  et~al., 2017, \mn@doi [MNRAS] {10.1093/mnras/stx1338}, \href
  {https://ui.adsabs.harvard.edu/abs/2017MNRAS.470.3775V} {470, 3775}

\bibitem[\protect\citeauthoryear{{Wang}, {Koribalski}, {Serra}, {van der
  Hulst}, {Roychowdhury}, {Kamphuis}  \& {Chengalur}}{{Wang}
  et~al.}{2016}]{Wang:2016}
{Wang} J.,  {Koribalski} B.~S.,  {Serra} P.,  {van der Hulst} T.,
  {Roychowdhury} S.,  {Kamphuis} P.,   {Chengalur} J.~N.,  2016, \mn@doi
  [MNRAS] {10.1093/mnras/stw1099}, \href
  {https://ui.adsabs.harvard.edu/abs/2016MNRAS.460.2143W} {460, 2143}

\bibitem[\protect\citeauthoryear{{Wolfe}, {Gawiser}  \& {Prochaska}}{{Wolfe}
  et~al.}{2005}]{Wolfe:2005}
{Wolfe} A.~M.,  {Gawiser} E.,   {Prochaska} J.~X.,  2005, \mn@doi [ARA\&A]
  {10.1146/annurev.astro.42.053102.133950}, \href
  {http://adsabs.harvard.edu/abs/2005ARA%26A..43..861W} {43, 861}

\bibitem[\protect\citeauthoryear{{Wright}, {Griffith}, {Hunt}, {Troup}, {Burke}
   \& {Ekers}}{{Wright} et~al.}{1996}]{Wright:1996}
{Wright} A.~E.,  {Griffith} M.~R.,  {Hunt} A.~J.,  {Troup} E.,  {Burke} B.~F.,
   {Ekers} R.~D.,  1996, \mn@doi [ApJS] {10.1086/192272}, \href
  {http://adsabs.harvard.edu/abs/1996ApJS..103..145W} {103, 145}

\bibitem[\protect\citeauthoryear{{Zafar}, {P{\'e}roux}, {Popping}, {Milliard},
  {Deharveng}  \& {Frank}}{{Zafar} et~al.}{2013}]{Zafar:2013}
{Zafar} T.,  {P{\'e}roux} C.,  {Popping} A.,  {Milliard} B.,  {Deharveng}
  J.-M.,   {Frank} S.,  2013, \mn@doi [A\&A] {10.1051/0004-6361/201321154},
  \href {http://adsabs.harvard.edu/abs/2013A%26A...556A.141Z} {556, A141}

\bibitem[\protect\citeauthoryear{{Zwaan}, {van der Hulst}, {Briggs},
  {Verheijen}  \& {Ryan-Weber}}{{Zwaan} et~al.}{2005}]{Zwaan:2005}
{Zwaan} M.~A.,  {van der Hulst} J.~M.,  {Briggs} F.~H.,  {Verheijen} M.~A.~W.,
   {Ryan-Weber} E.~V.,  2005, \mn@doi [MNRAS]
  {10.1111/j.1365-2966.2005.09698.x}, \href
  {http://adsabs.harvard.edu/abs/2005MNRAS.364.1467Z} {364, 1467}

\makeatother
\end{thebibliography}

%%%%%%%%%%%%%%%%%%%%%%%%%%%%%%%%%%%%%%%%%%%%%%%%%%

%%%%%%%%%%%%%%%%% APPENDICES %%%%%%%%%%%%%%%%%%%%%

%%%%%%%%%%%%%%%%%%%%%%%%%%%%%%%%%%%%%%%%%%%%%%%%%%

% Don't change these lines
\bsp	% typesetting comment
\label{lastpage}
\end{document}